\documentclass[aps,prd,floatfix,superscriptaddress,twocolumn,showpacs,amssymb]{revtex4-1}

\usepackage{epsfig}
\usepackage{graphicx}
\usepackage{dcolumn}
\usepackage{bm}
\usepackage{ctable}

\newcommand{\be}{\begin{equation}}
\newcommand{\ee}{\end{equation}}
\newcommand{\bea}{\begin{eqnarray}}
\newcommand{\eea}{\end{eqnarray}}
\newcommand{\hf} {\frac{1}{2}}
\newcommand{\nn}{\nonumber\\}

\def\eq#1{(\ref{#1})}
\def\la{\langle}
\def\ra{\rangle}

\def\ord#1{{\cal O}\left(#1\right)}

\def\t#1{{\tilde{#1}}}
\def\c#1{{\cal{#1}}}
\def\h#1{{\hat{#1}}}

\begin{document}
\title{Particle in a cavity in one-dimensional bandlimited quantum mechanics
}

\author{K. Sailer}
\affiliation{Department of Theoretical Physics, University of Debrecen,
P.O. Box 5, H-4010 Debrecen, Hungary}

\author{Z. P\'eli}
\affiliation{Department of Theoretical Physics, University of Debrecen,
P.O. Box 5, H-4010 Debrecen, Hungary}

\author{S. Nagy}
\affiliation{Department of Theoretical Physics, University of Debrecen,
P.O. Box 5, H-4010 Debrecen, Hungary}

\date{\today}

\begin{abstract}
The effects of the generalized uncertainty principle (GUP) on the low-energy stationary 
states of a particle  moving in a cavity with no sharp boundaries are determined by means
 of the perturbation expansion in the framework of one-dimensional bandlimited quantum
 mechanics. A realization of GUP resulting in the existence of a finite ultraviolet (UV)
 wave-vector cutoff $K\sim 1/\ell_P$ (with the Planck length $\ell_P$) is considered.
 The cavity of the size $\ell \gg \ell_P$ is represented by an infinitely deep trapezoid-well
 potential with boundaries smeared out in a range $R$ satisfying the inequalities
 $\ell\gg R\agt \ell_P$.  In order to determine the energy shifts of the low-lying stationary states,
  the usual perturbation expansion is reformulated in a manner that enables one to
 treat consistently order-by-order the direct and indirect GUP effects, i.e., those due to the
 modification of the Hamiltonian and the lack of the UV modes, respectively. It is shown
that the leading terms of the indirect and the direct GUP effects are of the first and second order,
 respectively, in the small parameter $\ell_P/\ell$ in agreement with our previous finding in a more 
naive approach  \cite{Sail2013}.

\end{abstract}



\maketitle
\section{Introduction}\label{intro}

Here our goal is to determine the stationary states of a particle in a cavity
in the framework of the one-dimensional bandlimited quantum mechanics.
This issue is a reconsideration of the `particle in the box' problem, but
now we remove the sharp boundaries of the cavity  describing it by
a trapezoid-well potential of infinite depth. Furthermore, we implement the nondegenerate
stationary perturbation expansion in a manner that enables us to distinguish of the various
 orders of the direct and indirect GUP effects. In this way we shall recover
our basic finding in  \cite{Sail2013} in a more reliable framework, namely, that
the indirect GUP effect is much more important than the direct GUP effect as to the energy
 shifts of the low-lying stationary states.

In the last two decades there has been a continuous interest in effective  quantum mechanics
 based on GUP \cite{Magg1994,Kempf199612,Bambi2008,Adler2010,Hossa2010,Noza2011,Piko2011,Hosse2012},
 the various modifications of Heisenberg's uncertainty principle, that is motivated by
 theoretical indications \cite{Mead1964,foam,string,Tiwar2008} pointing towards the
 consequences quantum gravity may have on the behaviour of low-energy quantum systems.   
 Among the various realizations of GUP there is a class when the deformation of the 
commutator 
relation for the operators of the coordinate $\h{x}$ and the canonical momentum $\h{p}_x$
 depends only on the canonical momentum,
\bea\label{gup}
  \lbrack \h{x}, \h{p}_x\rbrack &= &i\hbar f( \alpha |\h{p}_x|)
\eea
with   the deformation function $f(|u|)$, where $u=\alpha p_x$ and $\alpha =\ord{ a/\hbar}$
 with some minimal length scale $a$, a kind of the quantum of distance. Here we shall use the
 wide-spread conjecture that $a$ is equal to the Planck length $a=\ell_P\approx 1.616\times
 10^{-35}$ m, although an unquestionable proof for that does not exist  \cite{Hosse2012}. 
 Therefore there is an alternative view point that the minimal length scale $a$ must be
 determined from experimental data and various upper bounds on it have been given
\cite{Arka1998,Chang2001,Chang2002,Benc20024,Benc20029,Chang2004,Benc2005,Benc2007,Chang2011,
Lewi2011}. In the present paper we shall  restrict ourselves to the particular form of the 
 deformation function $f=1+\alpha^2 p_x^2$ for which there exists a  minimal wavelength, i.e.,
 a maximal magnitude $K= \pi/a$ of the wave vector,  but the canonical momentum 
$p_x = \alpha^{-1} \tan(\alpha \hbar k_x) $ can take arbitrarily large values
 \cite{Kempfma1994,Kempf199604}. Then the physical states are restricted to those of finite
 bandwidth, i.e., the wave-vector  operator $\h{k}_x=(\alpha\hbar)^{-1} {\rm{arctan}}(
\alpha \h{p}_x)$ can only take eigenvalues in the interval $\lbrack -K, K\rbrack$.

In the present paper we shall replace the infinitely deep square-well potential, generally 
used in  the discussions of the `particle in the box' problem, with a potential with smeared 
out boundaries. It was noticed already in \cite{Pedra201112} that the usage of a potential
 with sharp boundaries is in contradiction with the existence of the minimal length scale $a$
 in bandlimited quantum mechanics.  Space has  particular features  in one-dimensional 
bandlimited quantum mechanics \cite{Kempf1993}. Although the wave-vector and the canonical
 momentum operators are self-adjoint, the coordinate operator $\h{x}$ is only Hermitian 
symmetric. It has, however, a one-parameter family of self-adjoint extensions with
 eigenvalues determining a grid with equidistant spacing  on the coordinate axis, defining a 
minimal length scale $a$, while the grids belonging to the various extensions can be
 continuously shifted into each other. This is a sign that positions can only be observed
 with a maximal precision $\Delta x_{min}\approx a$, while  momenta can be measured with
 arbitrary accuracy \cite{Kempfma1994,Hinr1995,Kempf1996,Deto2002,
Bang2006,Dorsch2011, Noza2011, Mend2011}. There exist formal coordinate eigenstates,
 representing the eigenvectors of any of the self-adjoint extensions of the coordinate
 operator, but those   cannot be approximated now by a sequence of physical states with
 uncertainties in position  decreasing to zero \cite{Kempfma1994}. Nevertheless, they form
 an appropriate basis to decompose any physical state into a linear superposition of
 coordinate eigenstates. In our previous paper \cite{Sail2013} we discussed in detail that
 such a decomposition of any physical state corresponds to the generalization of Shannon's 
sampling theorem \cite{Shan49}. In bandlimited quantum mechanics the coordinate space turns
 out to exhibit features of discreteness \cite{Das2010} and continuity  at the same
 time like information does \cite{Kempf199709,Kempf199806,Kempf199905,Kempf2007,Kempf2010,
Kempf201010,Bojo2011}. The main idea is that space can be thought of a differentiable
 manifold, but the physical degrees of freedom cannot fill it in arbitrarily dense manner.
It has also been conjectured that degrees of freedom corresponding to structures smaller
 than the resolvable Planck scale turn into internal degrees of freedom 
\cite{Kempf199706,Kempf199810,Kempf199811,Kempf199907}. The square-integrable coordinate 
wavefunctions do not have the usual probabilistic meaning, yet provide  useful tools to
 characterize the quantum states of the particle which can then be analyzed e.g. in terms of
 maximally localized states \cite{Hinr1995,Kempf1996,Deto2002,Bang2006}.
 
Keeping in mind the above described structure of space, now we choose an infinitely deep 
trapezoid-well potential to model the cavity. Then the   boundaries of the cavity
are smeared out in intervals of the size $2R$, where $R$ is much less than
the size $\ell$ of the cavity. It is assumed that the inaccuracy  $\sim R$ of the
 determination of the positions of the boundaries  should be of the order
of the minimal length scale if that inaccuracy  is caused solely  by  the space structure
 at the Planckian length scale. In other terms it is assumed that no many-body physics of the
medium surrounding the cavity can provide boundaries more sharp than it is allowed by the 
existence of the minimal length scale. In realistic cases, when the cavity is physically prepared
in a medium and is of the size on nanoscales, the unsharpness of its  boundaries is basicly 
determined by some interactions of range much larger than the Planck scale, but even then can
hold the inequality $R\ll \ell$. The result obtained below shall be valid for any values of the
 parameter $R$ for which the inequalities $\ell \gg R \agt a$ hold.
The grid of the coordinate eigenvalues seems to dictate that the distances $\ell$ and $R$ have to
be considered as integer times the spacing $a$.

In order to determine the low-energy spectrum of the particle in the cavity we shall solve
  the Schr\"odinger equation for the coordinate wavefunctions in the case when GUP implies
 finite bandwidth. It is well-known that GUP directly affects the Hamiltonian through the 
modification of the canonical momentum and that of the kinetic energy operator
 $ (\h{p}_x^2/2m)-(\hbar^2 \h{k}_x^2/2m)$ which can be expanded -- when low-energy states are
 considered -- in the powers of the small parameter $\alpha k\sim a/\ell$ where $k$ is
 the characteristic wave vector of the given stationary state.  This direct GUP effect has
 been treated in the framework of the  perturbation expansion using the  coordinate 
representation and discussed in detail for various quantum systems (see  Refs.  
\cite{
Kempf199604,Ali2009,Ali2011,Nozaz2005,Das2008,Jana2009,Mirza2009,Noza2010,
Ali2010,Pedra2010,Spren2010,Dasma2011,Pedra201110,Pedra201204,Valta2012,
Tau2012,Noza200507,Pedra2011,Pedra201309,Boua201311,Boua201312,Parwa2013} without the quest of completeness), among others for the
 particle in a box \cite{Ali2009,Ali2010,Ali2011,Pedra2010,Pedra201110,Noza200507,Pedra2011,Blado2014}. 
In that connection it was discussed the problem of prescribing boundary conditions
 \cite{Pedra2013,Louko2014} and the existence of the self-adjoint extensions of various forms of the GUP
 modified Hamiltonian  \cite{Bala2014}. Generally the perturbative discussions of the various
low-energy quantum systems  were performed  either for realizations of GUP when no wave-vector cutoff $K$
 occurs or with the neglection of the   indirect GUP effect caused by such an UV wave-vector cutoff.
One were inclined to think that the existence of a very high UV cutoff, $K\sim\pi/a\sim 10^{35}$ m$^{-1}$
has negligible effect on the low-energy states of the quantum system. Nevertheless, in our previous paper
 \cite{Sail2013} it was shown 
 on the example of  the `particle in the box' problem that the indirect GUP effect may be even much more
important than the direct GUP effect is.  The energy shifts of the low-energy stationary states of a
 particle in the infinitely deep square-well potential were determined 
in the framework of the bandlimited quantum mechanics considering the latter as an effective
 theory in which no quantum fluctuations of wavelength smaller than those
 of the order of the minimal length are possible, i.e., in which the coordinate wavefunctions
 should not contain Fourier components with wave vectors outside of the finite band 
 $k_x\in \lbrack -K, K\rbrack$. In order to built in this restriction into the Schr\"odinger
 equation we developed a projection technique that essentially means that the wavefunctions
 and the operators are projected by means of the projector $\h{\Pi}$  onto the  bandlimited
 subspace $\c{H}$ of the Hilbert space. We have seen that in the case of the stationary
 states of the particle in the box, i.e., in a square-well potential of infinite depth the 
indirect GUP effect resulting form the existence of the finite bandwidth is an effect of
 the first order, while for a nonrelativistic particle the direct GUP effect is an effect of 
the second order in the small parameter $a/\ell$. Motivated by this finding we shall develop
 here a perturbation expansion up to the second order making the working hypothesis that the
 indirect and the direct GUP effects are of the first and second orders, respectively. We
 shall show that  the consistent separation of the various orders in the small parameter
$a/\ell$ becomes possible  in the case of infinitely deep  trapezoid-well potential that
justifies {\em a posteriori} our working hypothesis. In this manner we can confirm our
 previous result in a more reliable framework.
 
A similar result, but in somewhat other terms was found for the particle in a box in Ref.
 \cite{Pedra201112} where it was realized  that the model with precisely given box size
 $\ell$ is ill-defined in the sense that a change 
of the box size of the order $\ell_P$, i.e., that of the maximal accuracy $\Delta x_{min}$ of
 the position determination causes an energy shift of the order $\ord{\ell_P/\ell}$ as
 compared to the direct GUP effect of the order $\ord{(\ell_P/\ell)^2}$. Here we improve
 our previous approach to the `particle in box' problem in \cite{Sail2013} by (i) replacing 
the box, the infinitely deep square-well potential with  sharp boundaries
 by an infinitely deep trapezoid-well potential with  smeared out boundaries (ii) and
 by an order-by-order consistent usage of the perturbation expansion. The main message, the 
dominance of the indirect GUP effect over the direct one has been recovered. The sharp 
boundaries of the box potential caused periodic dependence of the relative energy shifts
of the low-energy stationary states on the ratio $\ell/a$ \cite{Sail2013}. Now 
we  shall see that a similar periodic dependence on the ratio $(\ell-2R)/a$ remains  still present
 after the smearing out the boundaries of the cavity in the range $2R$ when the 
trapezoid-well potential is used, since the `effective length' of the cavity equals now to
 $\ell-2R$.  

Our  paper is constructed as follows. In Section \ref{perte} the perturbation Hamiltonian  
 and the usage of the nondegenerate stationary perturbation expansion are given. In Section \ref{pincav}
we apply the perturbation expansion to the determination of the stationary states of the particle
in the cavity modelled by an infinitely deep trapezoid-well potential and discuss the results
obtained for the relative energy shifts of the low-energy states due to the GUP effect.
A short summary is given in Section \ref{sum}. In Appendix \ref{unpert}  
the unperturbed Schr\"odinger equation is solved for the particle in the trapezoid-well potential
keeping the depth of the potential finite but much larger than the particle's energy.
A reminder for the derivation of the explicit formulas of  the nondegenerate stationary perturbation
 expansion is given in Appendix \ref{perexp} for the case when the perturbation Hamiltonian itself
consists of various order terms. Finally, the evaluation and estimation of the matrix elements of the
various terms of the perturbation Hamiltonian are given in Appendix \ref{matele}.

\section{Perturbation expansion}\label{perte}

We describe the cavity by the trapezoid-well potential parametrized as 
 \bea\label{trapez} 
\lefteqn{
  V_t(x)      }\nn
 &= & V_\infty \Biggl\{\begin{array}{lll}
1  &{\rm{for}}& (\ell/2)+R\le  |x|,\cr
  \frac{R+|x|-(\ell/2)}{2R}  &{\rm{for}}& (\ell/2)-R\le |x| <(\ell/2)+R,\cr
  0  &{\rm{for}}& 0\le  |x| <(\ell/2)-R .\end{array}\nn
\eea
The potential $V_t(x)$ is symmetric, $V_t(x)=V_t(-x)$,  takes the constant value $V_\infty$  
asymptotically far away from the origin, and decreases to and increases from zero linearly
at the boundaries of the cavity at $x=-\ell/2$ and $\ell/2$, respectively, in an interval of
the length $2R\ll \ell$. In this way the boundaries of the cavity are smeared out on a distance $2R$.
In the limit $V_\infty\to\infty$ the potential becomes infinitely deep and formally it goes over
 into a square-well potential of the size $\ell-2R$. The limit is, however, not smooth, because
for any finite but large values of $V_\infty$ the boundary regions of the potential-well still keep
their size. Therefore, the quantum mechanical expectation values of some observables of the
particle in stationary states may become different for an infinitely deep trapezoid-well potential
and for an infinitely deep square-well potential. We shall perform the calculations for
a trapezoid-well potential with finite but large values of $V_\infty$ and  take the limit
$V_\infty\to \infty$ when the energy eigenvalues have already been evaluated. We shall see that the
results obtained for the relative energy shifts of the stationary states differ somewhat from those
 obtained in our previous work \cite{Sail2013}. One has to notice that a potential imitating
a cavity of the size $\approx \ell$ with obscure boundaries can be chosen in infinitely many ways.
Our choice is simple enough enabling us to find analytic solution of the unperturbed Schr\"odinger
 equation.

In the bandlimited quantum mechanics the stationary Schr\"odinger-equation,
\bea\label{stsch}
  \h{H}\psi_\nu &=&\varepsilon_\nu \psi_\nu
\eea
 for the particle of mass $m$ in the cavity  represented by the trapezoid-well potential  \eq{trapez}
is given in terms of the Hamiltonian with the kernel
\bea
  H(x,y) &=&T_\Pi(x,y) + V_\Pi (x,y),
\eea
where
\bea
 T_\Pi(x,y) &=& \int_{-\infty}^\infty dz \Pi( x-z) \frac{ \lbrack \tan (-i\alpha \hbar \partial_z ) \rbrack^2}{2m\alpha^2}  \Pi(z-y)\nn
\eea
and
\bea
V_{\Pi}(x,y)&=& \int_{-\infty}^\infty dz\Pi(x-z)V_t(z)\Pi(z-y)
\eea
are the kernels of the projected kinetic and potential energy operators, respectively, and the
kernel of the projector $\h{\Pi}$  is given as
\bea\label{projec}
  \Pi(x-y)=  \int_{-K}^K \frac{dk_x}{2\pi} e^{ik_x(x-y)}= \frac{\sin \lbrack K(x-y)\rbrack}{\pi 
(x-y)}
\eea
(see \cite{Sail2013}). Instead of solving the stationary Schr\"odinger-equation \eq{stsch} directly, we
 shall determine the stationary states by means of the perturbation expansion. The Hamiltonian kernel of
 the unperturbed system is
 now chosen as
\bea\label{hunpt}
  H_0(x,y) &=& \biggl( -\frac{\hbar^2}{2m} \partial_x^2 + V_t(x) \biggr)\delta(x-y).
\eea
Then the solution of the stationary Schr\"odinger-equation,
\bea\label{stschup} 
  \h{H}_0 \phi_\nu &=& \epsilon_\nu \phi_\nu,
\eea
 i.e.,  the determination of the stationary states of the unperturbed
system is a problem stated in the framework of ordinary quantum mechanics and can be solved
 analytically  (see Appendix \ref{unpert}). The choice \eq{hunpt} of the unperturbed Hamiltonian 
reflects our working hypothesis that both the direct GUP effect, i.e., the effect due to the 
modification of the kinetic energy operator, as well as the indirect GUP effect occurring due to
 the projection to the subspace of bandlimited wavefunctions can be considered as perturbations.
 Such an assumption is motivated by previous findings for the low-energy stationary states of a
 nonrelativistic particle in a box, i.e., in an infinitely deep square-well potential exhibiting
 sharp boundaries. On the one hand, it is well-known that the direct GUP effect is of the order
 $(a/\ell)^2$ for a nonrelativistic particle in a box
 \cite{Noza200507,Ali2009,Ali2010,Pedra2010,Pedra2011,Pedra201110,Ali2011}
and, on the other hand, we have found  that the indirect GUP effect is even more important but
 still perturbative and of the order $a/\ell$ \cite{Sail2013}. In the latter case we did not, 
however,  made a distinction between the contributions of the various orders in the small 
parameter $a/\ell$. Our purpose is now to show in the framework of a more consistent perturbative 
treatment that the indirect GUP effect caused by the lack of UV wave-vector modes is perturbative
 and of the order $a/\ell$ indeed. Moreover, with the choice of the trapezoid-well potential of 
infinite depth instead of the square-well potential with sharp boundaries, we would like to remove
the inconsistency pointed out in \cite{Pedra201112}.

 Ordinary quantum mechanics for a particle in a box of size $\ell$ can be used to estimate the 
 low-energy part of the spectrum as $\epsilon_\nu^{box} =\frac{\hbar^2\pi^2 (\nu+1)^2}{2m\ell^2}=
\frac{\hbar (k_\nu^{box})^2}{2m}$ with the wave vector $k_\nu^{box}=\frac{\pi (\nu+1)}{\ell}$
with $\nu=0,1,2,\ldots$. We expect that the energies $\epsilon_\nu$ and $\varepsilon_\nu$ are
 close to the values $\epsilon_\nu^{box}$ in the low-energy part of the spectrum when the depth of
 the potential well $V_\infty$ is chosen to be much larger than $\epsilon_0^{box}$. Finally, we shall
 take the limit $V_0\to \infty$ and that enables us to separate the various orders of the
 perturbation expansion in a rather straightforward manner. 

As compared to the unperturbed Hamiltonian $\h{H}_0$ the direct GUP effect results in the 
deviation  $\h{h}_t=(\h{p}_x^2-\hbar^2\h{k}_x^2 )/(2m)$ of the GUP modified kinetic energy operator
from the ordinary  one. The indirect GUP effects arise from the modification $\h{t}=\h{T}_\Pi-\h{T}$
 and $\h{v}=\h{V}_{t\Pi}-\h{V}_t$  of the kinetic and potential energy operators, respectively due 
to projection onto the bandlimited subspace. Then the  Hamiltonian of the perturbation, 
$\delta \h{H}=\h{H}-\h{H}_0$ can be rewritten as 
\bea
  \delta \h{H}&=& \h{h}_t + \h{t} +\h{v}
\eea
with the kernels
\bea\label{kerns}
  h_t(x,y) &=& \frac{1}{2m}\biggl(
 \alpha^{-2} \tan^2(-i\alpha\hbar \partial_x)
 - \hbar^2 \partial_x^2\biggr)
\delta(x-y),\nn
 t(x,y) &=& \frac{1}{2m\alpha^2}  \tan^2 (-i\alpha\hbar \partial_x) \lbrack \Pi(x-y)-\delta(x-y)\rbrack,\nn
 v(x,y) &=& \hf \lbrack V_{t}(x)+V_{t}(y)\rbrack \lbrack \Pi(x-y)
- \delta (x-y)\rbrack,
\eea
where we applied  the rules established in \cite{Sail2013} to write down the kernels of the
projected operators in the framework of bandlimited quantum mechanics. The usual nondegenerate
 stationary perturbation expansion provides the corrections of the various orders to the energy levels
 and the corresponding wavefunctions in terms of the matrix elements
 $\la \phi_\nu|\delta \h{H} |\phi_{\nu'}\ra$ of the perturbation Hamiltonian $ \delta \h{H}$
 sandwiched by the unperturbed stationary wavefunctions $\phi_\nu$ and $\phi_{\nu'}$. Now we shall apply
 this perturbation scheme assuming that the matrix elements of the perturbation Hamiltonian can be
 recasted into the sum of those of perturbation operators $\delta^{[n]}\h{H}$ of various orders $n$
in the small parameter $a/\ell$, for which  
\bea\label{perham}
  \delta \h{H} &=& \sum_{n=1}^\infty \delta^{[n]}\h{H}
\eea
holds. We shall show that in the limit $V_0\to \infty$ it is possible to identify the  pieces
of various orders of the  matrix elements $\la \phi_\nu|\delta \h{H} |\phi_{\nu'}\ra$, 
i.e., those of the matrix elements of the operators $\h{h}_t$, $\h{t}$, and $\h{v}$.

As compared to the unperturbed Hamiltonian $\h{H}_0$ the direct GUP effect results in the 
additional terms $\h{h}_t=\sum_{n=1}^\infty \h{h}^{(n)}_t$  modifying the ordinary kinetic energy
 operator, where $\h{h}^{(n)}_t \sim \ord{ (a/\ell)^n}$.
 For the   deformation function  $f(p_x) = 1+\alpha^2 p^2_x$ the expansions
  $p_x =\tan u\approx u+ \frac{u^3}{3}+\ldots$ with $u=\hbar k_x$ and
$ \tan^2 u\approx u^2+ \frac{2u^4}{3}= u^2\biggl( 1+\frac{2}{3} u^2\biggr)$
imply the kernels $h_t^{[1]}(x,y)=0$ and
\bea\label{ht2}
  h_t^{[2]}(x,y) &=& \frac{2\alpha^2}{3} \frac{\hbar^4}{2m} \partial_x^4\delta (x-y).
\eea
Similarly the piece $\h{t}$ of the perturbation  Hamiltonian  associated with the projection of 
the kinetic energy operator can be split as $\h{t}=\h{t}^{[1]}+\h{t}^{[2]}+\ldots$ with the
kernels
\bea\label{t1}
  t^{[1]}(x,y) &=& -\frac{\hbar^2}{2m} \partial_x^2 \lbrack \Pi(x-y)-\delta(x-y)\rbrack,\\
\label{t2}
 t^{[2]}(x,y) &=& \frac{2\alpha^2}{3} \frac{\hbar^4}{2m} \partial_x^4\lbrack \Pi(x-y)-\delta(x-y)\rbrack,
\eea
where the presence of the projector $\h{1}-\h{\Pi}$ onto the subspace of the UV modes is treated
as the one causing a first-order effect. In the same line also the perturbation $\h{v}$ with
the kernel
\bea\label{v1}
  v^{[1]}(x,y) &=& \hf \lbrack V_{t}(x)+ V_{t}(y) \rbrack \lbrack \Pi(x-y)-\delta(x-y)\rbrack
\eea
shall be treated as that of the first order.
At the end of the evaluation of the matrix elements
of these operators between the unperturbed states $\phi_\nu$ we shall see that they contain terms of 
higher orders too. This seeming inconsistency can however be cured by reshuffling the appropriate
second order terms from the matrix elements $\la \phi_\nu |\delta \h{H}^{[1]}|\phi_{\nu'}\ra$ into 
the matrix elements $\la \phi_\nu |\delta \h{H}^{[2]}|\phi_{\nu'}\ra$ and neglecting
all contributions of higher than second order when we restrict ourselves to the second order of 
the perturbation expansion, at which the direct GUP effect comes into play. Therefore, the selection
 of the terms of various orders made above on the level of operators  has to be reconsidered and made
 more accurate at the level of the matrix elements. This happens because the projector introduces
$a$-dependence into  the kernels of the operators that cannot be Taylor-expanded at the operator
 level. We shall see that for the trapezoid-well potential of infinite depth the separation of
the various orders at the level of the matrix elements is straightforward.
It should be noted that the possibility to treat the indirect GUP effect as perturbation
seems to be a particular feature  of the case when the particle moves in a potential of infinite depth.
 For a potential of finite depth there would occur nonanalytic  dependences of 
the matrix elements on the length scale $a$ through the explicit dependence of the projector
\eq{projec} on the wave-vector cutoff $K=\pi/a$ (see e.g. our remark at the end of the paragraph following
Eq. \eq{t1_2_2limit} in Appendix \ref{matele}).  This makes questionable the perturbative treatment
of the indirect GUP effect in the  case of a  potential well of finite depth.

Now assuming that we keep the  terms of various orders of the perturbation Hamiltonian
under control, we insert its expansion \eq{perham} together with the expansion of the energy eigenvalue
and that of the wavefunction,
\bea\label{expas}
 \varepsilon_\nu &=& \epsilon_\nu + \sum_{n=1}^\infty \delta^{[n]}\epsilon_\nu,\nn
  \psi_\nu &=& \phi_\nu + \sum_{n=1}^\infty \delta^{[n]} \psi_\nu,
\eea
respectively, into the stationary Schr\"odinger equation \eq{stsch} and repeat the  steps being a piece 
of textbook material on nondegenerate perturbation expansion in ordinary quantum mechanics (for a
 reminder see Appendix \ref{perexp}). This yields then for the corrections of the first order
\bea\label{epsfirst}
  \delta^{[1]}\epsilon_\nu &=& \la \phi_\nu |\delta^{[1]} \h{H} | \phi_\nu\ra,
\\
\label{psifirst}
\delta^{[1]} \psi_\nu &=& 
- \sum_{\mu\not= \nu} \frac{ \la \phi_\mu | \delta^{[1]}\h{H}| \phi_\nu \ra}{ \epsilon_\mu-\epsilon_\nu}
\phi_\mu,
\eea
and those of the second order,
\bea\label{epssec}
  \delta^{[2]}\epsilon_\nu &=& \la \phi_\nu |\delta^{[2]} \h{H} | \phi_\nu\ra
- \sum_{\mu\not=\nu}  \frac{ |\la \phi_{\mu} | \delta^{[1]}\h{H}| \phi_\nu\ra |^2 }{ \epsilon_\mu
-\epsilon_\nu}  ,\\
\label{psisec}
\delta^{[2]} \psi_\nu &=& 
- \sum_{\mu\not= \nu}\frac{ \la \phi_\mu |\delta^{[2]}\h{H} |\phi_\nu\ra }{ \epsilon_\mu
-\epsilon_\nu }\phi_\mu\nn
&&
+ \sum_{\mu,\mu'\not= \nu} \frac{ \la \phi_\mu| \delta^{[1]}\h{H}| \phi_{\mu'} \ra
 \la \phi_{\mu'} | \delta^{[1]}\h{H}| \phi_\nu\ra }{(\epsilon_\mu-\epsilon_\nu)
 (\epsilon_{\mu'}-\epsilon_\nu ) }\phi_\mu\nn
&&
-\sum_{\mu\not= \nu}\frac{  \la \phi_\nu |\delta^{[1]}\h{H}|\phi_\nu\ra\la \phi_\mu | \delta^{[1]}\h{H}| \phi_\nu\ra}{(\epsilon_\mu-\epsilon_\nu)^2 }\phi_\mu \nn
&&-\hf \sum_{\mu\not= \nu} \frac{ |\la \phi_\mu|\delta^{[1]}\h{H} |
\phi_\nu\ra |^2 }{ (\epsilon_\mu-\epsilon_\nu)^2 } \phi_\mu.
\eea 
 The real question is whether the perturbation expansion does work for the particle in the
trapezoid-well potential of infinite depth in the framework of the bandlimited quantum mechanics.
This essentially depends on whether the matrix elements
 $\la \phi_\nu|\delta^{[n]} \h{H}|\phi_{\nu'}\ra$ do survive the limit $V_\infty\to \infty$ and are of
the order $(a/\ell)^n$. This will be shown in the next section.

It should be noticed that the formulas in Eqs. \eq{epsfirst}-\eq{psisec} are valid for discrete
 spectrum, but we consider only the low-lying states with energies $\epsilon_\nu\ll V_\infty$
 when the contribution of the continuous spectrum can be neglected and it disappears completely in 
the limit $V_\infty\to \infty$ that we have taken finally.

\section{GUP effect on a particle in the cavity}\label{pincav}

In this section we investigate how GUP affects  the low-lying stationary states of the particle
 in the cavity represented by the trapezoid-well potential of infinite depth in the framework of
 the perturbation scheme outlined in the previous section, restricting ourselves to the second order
 of the perturbation expansion in the small parameter $a/\ell$. For this purpose, one needs to
 evaluate the matrix elements $\la \phi_\nu |\delta^{[n]} \h{H} |\phi_{\nu'}\ra$ for $n=1,2$ of the
perturbations in the basis of the unperturbed wavefunctions $\phi_\nu$.  The determination of the
wavefunctions $\phi_\nu$ of the bound stationary states in the trapezoid-well potential is an 
analytically solvable problem in ordinary quantum mechanics given in Appendix \ref{unpert} in detail 
for the case when $V_\infty$ is asymptotically large. In order to solve the stationary Schr\"odinger
 equation \eq{stschup} for a particle in the potential $V_t(x)$, one has to divide the real axis into 
the intervals $I_i$ with $i=I,II,III,IV,V$ given as
\bea\label{intervs}
  && I_I=(-\infty, -b-R\rbrack,~~I_{II}=\lbrack  -b -R, -b +R\rbrack,\nn
&&I_{III}= \lbrack  -b +R, b -R\rbrack,\nn
&& I_{IV}= \lbrack  b -R, b +R\rbrack, ~~I_{V}=\lbrack  b +R,\infty)
\eea
with boundaries at the points of discontinuities of the first derivative of the potential $V_t(x)$
and $b=\ell/2$. In Appendix \ref{unpert} we have followed the usual procedure of making an ansatz
 for the pieces of the wavefunction in the various intervals, then matching the neighbouring pieces
 by the conditions of continuity of the wavefunction and its first derivative, and finally normalizing
 the wavefunction. After a straightforward but lengthy calculation the following expression is found
 for the wavefunction $\phi_\nu$ for asymptotically large values  of the depth $V_\infty$ of the
 potential well,
\bea\label{finupieces}
  \phi_{\nu I}(x) &\sim &  \hf \pi^{-1/2}(\gamma 2R)^{-1/4}  e^{-\zeta} B e^{ \kappa_\nu (x- (-b-R))},\nn
  \phi_{\nu II}(x) &\sim& B {\rm{Ai}}( \gamma (-x -(b-R) -r_\nu)),\nn
 \phi_{\nu III}(x)&\sim&  \frac{ (-i)^\nu}{ 2 \sqrt{b-R} } \lbrack e^{ik_\nu x} 
+(-1)^\nu e^{-ik_\nu x}\rbrack,
\eea
where $\phi_{\nu i}(x)$ denote the pieces of the wavefunction in the intervals $I_i$ and the other pieces
 are given by the parity relations $\phi_\nu(x)= (-1)^\nu \phi_\nu(-x)$ due to the symmetry of the
 potential $V_t(-x)=V_t(x)$. Here ${\rm{Ai}}(x)$ stands for the Airy function falling down asymptotically
for $x\to \infty$,  and the other notations are
\bea
  B &\sim &  \frac{ k_\nu^\infty  }{ \gamma c_2 \sqrt{b-R} } ,
\eea
$k_\nu^\infty=\pi(\nu+1)/(\ell-2R)$, $ k_\nu=\sqrt{2m\epsilon_\nu}/\hbar$, $\kappa_\nu = 
\sqrt{ 2m (V_\infty -\epsilon_\nu)}/\hbar$, $\gamma= (mV_\infty)^{1/3}(\hbar^2 R)^{-1/3}$,
 $r_\nu= 2R\epsilon_\nu/V_\infty$, $R_\nu=2R-r_\nu$, and $\zeta=(2/3) (\gamma 2R)^{3/2}$, and
 $c_2=3^{-1/3}/\Gamma(1/3)\approx 0.259$ as given in Appendix \ref{unpert}. It is easy to notice 
that the wavefunctions \eq{finupieces} are real both for $\nu$ even and odd.
 
 It is worthwhile mentioning that in the limit $V_\infty\to \infty$, i.e., in the limit of the infinite
 potential depth  the pieces $\phi_{\nu I}(x)$ and $\phi_{\nu V}(x)$ of the wavefunctions are
 suppressed exponentially, due to their constant coefficients, the pieces $\phi_{\nu II}(x) $ and
 $\phi_{\nu IV}(x)$ are suppressed  due to their constant coefficients  by the factor
 $1/\gamma\sim V_\infty^{-1/3}$ and only  the pieces $\phi_{\nu III}(x)$ exhibit  coefficients of the order
 $V_\infty^0=1$.  As to the pieces  $\phi_{\nu II}(x) $ and $\phi_{\nu IV}(x)$, one has to be aware of the
 fact that those are the wavefunctions in the rather short intervals $I_{II}$ and $I_{IV}$, so that the
 arguments of the Airy functions are close to $\gamma(R-r_\nu)$ that takes asymptotically large positive
 values for $V_\infty\to \infty$. Consequently, the wavefunctions  are suppressed exponentially by the factor 
$e^{-\zeta}$ in the intervals $I_{II}$ and $I_{IV}$ rather than by the negative power $V_\infty^{-1/3}$ of
 $V_\infty$.  Therefore, the stationary wavefunctions in the infinitely deep trapezoid-well potential
 contract to the interval $I_{III}$, where the potential is vanishing, and  satisfy Dirichlet's boundary
 condition at $x=\pm  (b-R)$. In that limit they become identical to the wavefunctions of the stationary
 states in a box (in a square-well potential of infinite depth) of size $\ell-2R$.

Let us turn now to the determination of the matrix elements
 $\la \phi_\nu |\delta^{[n]} \h{H} |\phi_{\nu'}\ra$ for $n=1,2$.  Before going into any details, it is in
order to make a remark on our expectations how those should behave in the limit $V_\infty\to \infty$. The 
wavefunction of any state $\nu$  goes essentially over into the wavefunction of the state with the same
 quantum number $\nu$ in the box (in the square-well potential of infinite depth) of  the size $\ell-2R$.
 Therefore, we expect that the matrix elements will contain the contributions obtained for the square-well
 potential, but with the change of the size of the box from $\ell$ to $\ell-2R$. There may occur however
 modifications because the size $2R$ of the intervals $I_{II}$ and $I_{IV}$ is kept unaltered during  the 
limiting process $V_\infty\to \infty$. These intervals, where the boundaries of the cavity are smeared out
 have no analogues in the case of the square-well potential, and may alter the matrix elements. Therefore,
 one cannot say a priori how the matrix elements look like, and their evaluation is then unavoidable.

The matrix elements of any Hermitian symmetric kernel $K(x,y)=K^*(y,x)$ can be rewritten as the sum
\bea
  \int_{-\infty}^\infty dx  \int_{-\infty}^\infty dy  \phi_{\nu'}^* (x)K(x,y)
\phi_\nu(y) &=&
  \sum_{i,j=I}^{V} K^{ij}_{\nu'\nu}
\eea
with 
\bea\label{kijs}
  K^{ij}_{\nu'\nu}&=& \int_{I_i} dx\int_{I_j} dy   \phi_{\nu' i}^* (x)K(x,y)
\phi_{\nu j}(y).
\eea
These integrals are not independent. The kernels we have to do with are real symmetric ones, 
$K(x,y)=K(y,x)$ and the pieces of the wavefunctions are also real, that implies the relation 
$K^{ij}_{\nu'\nu}= K_{\nu\nu'}^{ji}$. Therefore we have to determine the integrals \eq{kijs} only 
for $(ij)=(I,I)$, $(I,II)$, $(I,III)$, $(II,II)$, $(II, III)$, $(III,III)$ for an arbitrary  choice
 $(\nu' \nu)$ of the pair of stationary states.

Let us consider the behaviour of the integrals  \eq{kijs} in the limit $V_\infty\to \infty$ for the
 various kernels in Eq. \eq{kerns}. After factorizing out the coefficients $A, B, D$ of the functions
  $\phi_{\nu I}(x)$ and $\phi_{\nu III}(x)$, their remaining terms   exhibit $V_\infty$-dependence only 
through $k_\nu$  that takes the finite limit $k_\nu^\infty$. Furthermore, the kernels $h_t(x,y)$ and
 $t(x,y)$ are independent of the parameter $V_\infty$. Therefore the integrals $(h_t)^{I,I}_{\nu'\nu}$,
 $(h_t)^{I,III}_{\nu'\nu}$, $t^{I,I}_{\nu'\nu}$, and $t^{I,III}_{\nu'\nu}$ vanish in the limit $V_\infty\to \infty$. 
In the integrals  $v^{I,I}_{\nu'\nu}$ and $v^{I,III}_{\nu'\nu}$ the kernel $v(x,y)$ contains an additional 
factor of $V_\infty$ in both or one of its terms, respectively. This, however, cannot compensate the
 exponential suppression caused by the factor $e^{-\zeta}$ in $\phi_{\nu I}(x)$, so that these integrals
 vanish in the limit $V_\infty\to \infty$ too. There is another trivial integral among the
 $K^{ij}_{\nu'\nu}$'s. Namely, $v^{III,III}_{\nu'\nu}=0$ vanishes identically, because the potential $V_t(x)$ 
vanishes in the interval $I_{III}$. It  should also be noted that the integrals where either one or both
 of the integration variables are from the interval $I_{II}$ contain the Airy-function in their integrands
 with a factor $\gamma\sim V_\infty^{1/3}$ in its argument, so that the integral should be taken first for 
finite but large $V_\infty$ in order to determine the limit $V_\infty\to \infty$. The integrals
 $K^{ij}_{\nu'\nu}$ not vanishing trivially in the limit $V_\infty\to \infty$ are then
 $(v^{[1]})^{I,II}_{\nu'\nu}$,  $(v^{[1]})^{II,III}_{\nu'\nu}$,  $(v^{[1]})^{II,II}_{\nu'\nu}$,
 $(t^{[1]})^{III,III}_{\nu'\nu}$, $(t^{[1]})^{I,II}_{\nu'\nu}$, $(t^{[1]})^{II,III}_{\nu'\nu}$, and
 $(t^{[1]})^{II,II}_{\nu'\nu}$  in the first order and 
 $(h_t^{[2]}+t^{[2]})^{III,III}_{\nu'\nu}$, $(h_t^{[2]}+t^{[2]})^{I,II}_{\nu'\nu}$,
 $(h_t^{[2]}+t^{[2]})^{II,III}_{\nu'\nu}$, and $(h_t^{[2]}+t^{[2]})^{II,II}_{\nu'\nu}$ in the second order.
The determination of these integrals is given in Appendix \ref{matele} and can be summarized as follows.

 The main conclusion of Appendix \ref{matele} is that  in the limit $V_\infty\to \infty$ all of the
 matrix elements we need in the perturbation scheme reduce to a single  contribution coming from
 the interval $I_{III}$ where the potential $V_{t}$ vanishes,
\bea
&&  (v^{[1]})_{\nu'\nu} \sim (v^{[1]})_{\nu'\nu}^{III,III},~~~~
   (t^{[1]})_{\nu'\nu} \sim (t^{[1]})_{\nu'\nu}^{III,III},\nn
&&
   (h_t^{[2]}+t^{[2]})_{\nu'\nu}\sim (h_t^{[2]}+t^{[2]})_{\nu'\nu}^{III,III}.
\eea
Moreover, the indirect GUP effect coming from the projection of the potential vanishes
 $(v^{[1]})_{\nu'\nu} \sim 0$ trivially, because the potential $V_t(x)=0$ in the interval $ I_{III}$, 
and the other nonvanishing matrix elements responsible for the direct and indirect GUP effects on
 the kinetic energy operator are diagonal in the state indices,
\bea
   (t^{[1]})_{\nu'\nu} &\sim & \delta_{\nu'\nu}  \frac{\hbar^2(k^\infty_{\nu})^2}{2m}
 \lbrack \c{I}_K(0)-1\rbrack,
\nn
   (h_t^{[2]}+t^{[2]})_{\nu'\nu}&\sim & 
 \delta_{\nu'\nu} 
\frac{2(k_\nu^\infty a)^2}{3}\frac{\hbar^2(k^\infty_{\nu})^2 }{2m} \c{I}_K(0) ,
\eea
where the asymptotic expression
\bea\label{ik0as}
 \c{I}_K(0)&\approx & 1  - \frac{2}{\pi} \biggl(\frac{2 a \cos (K(b-R))}{\pi(\ell-2R) } 
  -  \frac{4a^2\sin (K(b-R)) }{ \pi^2(\ell-2R)^2}  \biggr) \nn
\eea
is reliable due to the large cutoff $K=\pi/a$.  In the matrix element $  (h_t^{[2]}+t^{[2]})_{\nu'\nu}$
we have to keep only the leading order term of $\ord{a^2}$ due to the direct GUP effect when 
restricting ourselves to the second order of the perturbation expansion. The projection of the
 ordinary kinetic energy operator results in a sum of first and second order terms (when one neglects
 the higher-order ones in the asymptotic expansion of $\c{I}_K(0)$). Therefore, collecting the first
 and the second order terms appropriately, we should write
\bea\label{mateleperh}
  \la \phi_{\nu'} | \delta^{[1]} H| \phi_\nu\ra &\sim & 
   - \delta_{\nu'\nu}  \frac{\hbar^2(k^\infty_{\nu})^2}{2m} 
\frac{ 4a \cos (K(b-R))}{\pi^2(\ell-2R) } ,\nn
\la \phi_{\nu'} |\delta^{[2]} H |\phi_\nu\ra &\approx & \delta_{\nu'\nu}
\frac{\hbar^2(k^\infty_{\nu})^2}{2m} \biggl( 
\frac{2(k_\nu^\infty a)^2}{3} \nn
&&~~~~~~~~~~~~ + \frac{8a^2\sin (K(b-R)) }{ \pi^3(\ell-2R)^2} 
\biggr)
\eea
for $\hbar \alpha=a$. Here the first-order matrix element occurs purely due to the indirect GUP effect,
while the first and second terms of the second-order matrix element occured due to the direct and
 indirect GUP effects, respectively. We see that the matrix elements \eq{mateleperh} of the perturbation
 Hamiltonian survive the limit increasing the depth $V_\infty$ of the trapezoid-well potential to infinity.
Furthermore, we also see that in that limit the orders of magnitude of the various terms of the matrix
 elements \eq{mateleperh} are clearly given by their either linear or quadratic dependence on the small 
parameter $a/\ell$. Therefore the indirect GUP effect can be treated as perturbation that justifies our working hypothesis. This feature is, however, specific for the case when the potential well has
 infinite depth. For finite depth of the trapezoid-well potential there appear nonvanishing contributions
 to the matrix elements of the perturbation Hamiltonian which have nonanalytic dependence on the 
length scale $a$ that is introduced by the projection.  This casts doubt on the treatment of the
 indirect GUP effect as perturbation when the depth of the  potential well is finite.

Making use of the formulas \eq{epsfirst} and \eq{epssec} of the stationary 
perturbation expansion  we find the corresponding relative energy corrections of
 the energy levels,
\bea\label{rel1}
  \frac{\delta^{[1]}\epsilon_\nu}{\epsilon_\nu} &=& 
\frac{ \la \phi_{\nu'} | \delta^{[1]} H| \phi_\nu\ra }{\epsilon_\nu}=
    - \frac{ 4 \cos (K(b-R))}{\pi^2 }\frac{a}{\ell-2R},\\
\label{rel2}
   \frac{\delta^{[2]}\epsilon_\nu}{\epsilon_\nu} &=&
 \frac{ \la \phi_{\nu'} | \delta^{[2]} H| \phi_\nu\ra }{\epsilon_\nu}\nn
&=&\biggl(
\frac{2\pi^2 (\nu+1)^2}{3}
+ \frac{8\sin (K(b-R)) }{ \pi^3}   \biggr)
\biggl(\frac{a}{\ell-2R}\biggr)^2.\nn
\eea
Here we made use of the  diagonal behaviour of the matrix elements in the state indices. The first order
 contribution to the relative energy shift, \eq{rel1} is of the order $\ord{ a/\ell}$ and occurs
 purely due to the indirect GUP effect, while the second order one, \eq{rel2} is of the order
 $\ord{(a/\ell)^2}$ and occurs partially due to the direct GUP effect (the first term in the 
right-hand side of Eq. \eq{rel2}) and partially due to the indirect one (the second term in the 
right-hand side of Eq. \eq{rel2}). The direct GUP effect results in a positive  relative energy 
shift that increases  proportionally to $(\nu+1)^2$, i.e., proportionally to the energy $\epsilon_\nu$
 of the state. The indirect GUP effect is identical for all low-lying stationary states (keeping terms
 up to second order), but depends  periodically on the product  $K(b-R)=\hf (\pi/a) (\ell-2R)$ and its
 sign alternates  both in the first and in the second orders. At this point  we have to remember that
 all the approximations made during the derivation of the asymptotic form of the unperturbed wavefunction
 \eq{finupieces} in Appendix \ref{unpert} and during the evaluation and estimation of the matrix
 elements for infinite potential depth in Appendix \ref{matele} remain valid even if the parameter $2R$
 is set as small as the magnitude of  the minimal length scale, $2R\sim a$. The latter  choice means
 that the boundaries of the trapezoid-well potential are smeared out in an interval of the size of
 the minimal length scale. As to the indirect GUP effect, the proper choice of the size of the boundary
 regions seems however not to be very important, because its periodic  behaviour depends rather on the
 effective size $\ell-2R$ of the cavity. Were the length of the cavity quantized as argued in the case of
 the square-well potential of infinite depth in Refs. \cite{Ali2009,Bala2014}, i.e., would it hold the
relation $\ell-2R ={\rm{integer}}\times 2a$, the indirect GUP effect of the first order would have
alternating sign when this length increases with the amount of a single `quantum' $2a$.
 One should also mention that the wavefunctions keep their unperturbed forms
 in both the first and second orders of the perturbation expansion due to the diagonality of the matrix
 elements $  \la \phi_{\nu'} | \delta^{[n]} H| \phi_\nu\ra \sim  \delta_{\nu'\nu}$ with $n=1,2$ in the limit
$V_\infty\to \infty$.

\begin{widetext}

\begin{center}

\begin{table}[h]
\caption{\label{compar}  Comparison of  the various GUP effects for
the trapezoid-well and the square-well potentials of infinite depth.}
\begin{ruledtabular} 
\begin{tabular}{|c|c|c|}
\hline
 & &\cr
Type of effect & Trapezoid-well potential & Square-well potential \cr
& &\cr
\hline
\hline
& &\cr
 $R_h $ & $\frac{2}{3} (c\pi)^2 (\frac{a}{\ell-2R})^2
 (\nu+1)^2$ &  $\frac{2}{3} (c\pi)^2 (\frac{a}{\ell})^2
 (\nu+1)^2$ \cr
& &\cr
\hline
& &\cr
$R_v$ &  0 & $\frac{2}{\pi^2} \frac{a}{\ell} \biggl(
  2 + \lbrack (\nu+1)^2 \pi^2 -2 \rbrack \cos (K\ell) \biggr) $ for even $\nu$\cr
     &     & $0$ for odd $\nu$ \cr 
& &\cr
\hline
& &\cr
$R_t$ & $   -\frac{4}{\pi^2}  \frac{ a}{\ell-2R } \cos (K(\ell-2R)/2) +\frac{8}{\pi^3} (\frac{a }{ \ell-2R})^2\sin (K(\ell-2R)/2)$ & 
$  - \frac{ 4}{\pi^2} \frac{a}{\ell}  \cos (K\ell/2)- \frac{8}{\pi^3} (  \frac{a}{\ell})^2 \sin  (K\ell/2)$ ~~~~~~~~~ \cr
& &\cr
\hline
\end{tabular}
\end{ruledtabular}
\end{table}
\end{center}   

\end{widetext}

In Table \ref{compar} we  compare the results obtained now with those
 found in our previous paper \cite{Sail2013} for the square-well potential of infinite depth.
For comparison  we have used the asymptotic expansion
\bea
  \c{I}_K(\ell/2)-1&=&\frac{2}{\pi}{\rm{Si}}(K\ell/2) -1\nn
&\sim& 
  -\frac{2}{\pi} \biggl( \frac{2a \cos (K\ell/2)}{\pi\ell} +
  \frac{4a^2\sin  (K\ell/2)}{\pi^2\ell^2} \biggr)\nn
\eea
in order to rewrite our previous result on $R_t$  in \cite{Sail2013}.  
We see in Table \ref{compar} that both  potentials yield essentially the same relative energy 
shift $R_h$ of the order $(a/\ell)^2$ due to the direct GUP effect. In the expression of $R_h$ 
 there figures  the length of the interval in which the  potential vanishes, i.e.,  $\ell-2R$
 and  $\ell$ for the trapezoid-well and the square-well potentials, respectively.
 The most important change is that no indirect GUP effect occurs due to the projection of the 
potential operator in the case of the trapezoid-well potential. This is the consequence of the
existence of the intervals $I_{II}$ and $I_{IV}$ of finite size $2R$ where the trapezoid-well
 potential is smeared out. The width of these intervals remain unaltered when the limit
 $V_\infty\to \infty$ is taken and consequently the limits of the matrix elements of the projected 
potential operator are not identical for the trapezoid-well and the square-well potentials.
 It is worthwhile mentioning that the usage of the 
trapezoid-well potential leads to  relative energy shifts $R_v=0$ for all low-lying
stationary states, independently of their parity, as opposed to the peculiar result for the
square-well potential. The indirect GUP effect occurring due to the projection 
of the kinetic energy operator causes quite similar relative energy shifts
$R_t$ for both kinds of potentials, containing terms of both first and  second
orders. The size of the interval in which the potential vanishes figures in the explicit
 formulas for $R_t$ again.  The smearing out of the boundaries of the cavity does not removed the 
oscillatory behaviour of the relative energy shift $R_t$.

\section{Summary}\label{sum}

Summarizing, we have reconsidered the `particle in the box' problem in the framework of one-dimensional
bandlimited quantum mechanics by (i) using the infinitely deep trapezoid-well potential with smeared out
boundaries, and (ii) applying the nondegenerate stationary perturbation expansion to the determination of
 the low-energy stationary states by keeping track of the contributions of various orders  to the direct
 and indirect GUP effect. In that manner we confirmed the qualitative result found in \cite{Sail2013} in 
a more reliable framework. Namely, it is shown that the relative energy shift of the low-energy stationary
states of the nonrelativistic particle in the infinitely deep trapezoid-well potential show up
a contribution of the order  $\ell_P/\ell$  caused by the indirect GUP effect, while the direct GUP effect
on the energy shift is of second order $(\ell_P/\ell)^2$, where $\ell_P$ and $\ell$ are the Planck length 
and the length of the cavity, respectively. We argued that perturbative treatment of the indirect GUP effect
may not be possible for the potential well of finite depth.

\section*{Acknowledgements}
S. Nagy acknowledges the financial support from the J\'anos Bolyai
Programme of the Hungarian Academy of Sciences.

\appendix

\section{Stationary states of a particle in a trapezoid-well potential in ordinary quantum 
mechanics}\label{unpert}

Here we present the solution of the stationary Schr\"odinger equation \eq{stschup} 
with the unperturbed Hamiltonian \eq{hunpt} for the
 particle in the trapezoid-well potential $V_t(x)$ given in Eq. \eq{trapez} for bound states
when the energy of the particle is much less than the depth $V_\infty$ of the potential well.
In the various intervals $I_i$ with $i=I,II,\ldots, V$ given in \eq{intervs} the Schr\"odinger
 equation can be satisfied by the ansatz
\bea
   \phi_{\nu I}(x) &=& A e^{\kappa_\nu x},\nn
   \phi_{\nu II}(x) &=& B {\rm{Ai}}( f_{II}(x)) + C {\rm{Bi}}(f_{II}(x)),\nn
   \phi_{\nu III} (x) &=& D e^{ik_\nu x} + Ee^{-ik_\nu x}, \nn
   \phi_{\nu IV}(x)&=& F  {\rm{Ai}}( f_{IV}(x)) + G {\rm{Bi}}(f_{IV}(x)),\nn
   \phi_{\nu V}(x) &=& H e^{-\kappa_\nu x}
\eea
where the coefficients denoted by capital letters from $A$ to $H$ are yet unknown and
\bea
&& k_\nu = \sqrt{ 2m\epsilon_\nu}/\hbar,
~~\kappa_\nu = \sqrt{ 2m (V_\infty -\epsilon_\nu)}/\hbar.
\eea
The ansatz  for $ \phi_{\nu I}(x)$ and  $\phi_{\nu V}(x)$ is chosen in such a way that 
the wavefunction $\phi_\nu(x)$ were normalizable, i.e.,  the exponential functions falling
 down for $|x|\to \infty$ were chosen. In the intervals $I_{II}$ and $I_{IV}$, where the potential
is linearly falling and rising, respectively, the  Schr\"odinger equation can be recasted into the form 
of the differential equation for the Airy functions.  Multiplying the Schr\"odinger
 equation in the interval $I_{II}$ by  $2m/\hbar^2$, one rewrites the equation as
\bea
   \partial_x^2\phi_{\nu II} -\gamma^3 (-x+ \beta)\phi_{\nu II}&=&0,
\eea
with  $\beta= -b+R-r_\nu$, $b=\ell/2$, $r_\nu=2R\epsilon_\nu/V_\infty$, and $\gamma= (mV_\infty)^{1/3}(\hbar^2 R)^{-1/3}$. 
Then one transforms the independent variable $x$ to $\xi=\gamma(-x+\beta)$  and obtains for the function
 $\t{\phi}_{\nu II}(\xi(x))\equiv \phi_{\nu II}(x)$ the equation
  \bea
 \partial_\xi^2 \t{\phi}_{\nu II} (\xi) - \xi \t{\phi}_{\nu II} (\xi)&=&0
\eea
with the independent solutions ${\rm{Ai}}(\xi)$ and ${\rm{Bi}}(\xi)$, which implies that
\bea
  f_{II}(x)&=&\xi(x)= \gamma ( -x+ \beta ).
\eea
In the interval $I_{IV}$ we get similar expressions for the transformation $x\to  \xi = \gamma 
( x+ \beta )$ and find
\bea
     f_{IV}(x)&=& \gamma ( x+ \beta)=f_{II}(-x).
\eea

Since the potential is symmetric $V(x)=V(-x)$, the wavefunctions have either positive or
 negative parities, namely $+,-,+,-\ldots$ in the increasing order of their energies starting 
from the ground state. When the integer $\nu =0,1,2,\ldots $ enumerates the states with increasing energy, the parity is given by $(-1)^\nu$, i.e., $\phi_\nu (x)= (-1)^\nu \phi_\nu (-x)$.
 This implies the following restrictions on the coefficients
\bea
 && H= (-1)^\nu A,~~~~ F=(-1)^\nu B,\nn
&& G= (-1)^\nu C, ~~~~E= (-1)^\nu D.
\eea
The remaining 4 independent coefficients have to be determined from the boundary conditions
ensuring the continuity of the wavefunction and that of its first derivative,
\bea\label{bc1}
&&  Ae^{-\kappa (b+R)} = B {\rm{Ai}}(f_{II}(-b-R))\nn
&&~~~+ C {\rm{Bi}}(f_{II}(-b-R)),
\\
\label{bc2}
 &&A \kappa e^{-\kappa (b+R)} =- \gamma \lbrack B  {\rm{Ai}}^\prime (f_{II}(-b-R))
\nn
   && ~~~+ C {\rm{Bi}}^\prime (f_{II}(-b-R)) \rbrack,
\\
\label{bc3}
&& B {\rm{Ai}}(f_{II}(-b+R))+ C {\rm{Bi}}(f_{II}(-b+R))  \nn
&&
  ~~~= D \lbrack e^{ik_\nu(-b+R) } + (-1)^\nu  e^{-ik_\nu(-b+R) } \rbrack,
\\
\label{bc4}
&&- \gamma\lbrack  B {\rm{Ai}}^\prime(f_{II}(-b+R))+ C {\rm{Bi}}^\prime (f_{II}(-b+R)) 
 \rbrack\nn
&&~~~= ik_\nu  D \lbrack e^{ik_\nu(-b+R)  } - (-1)^\nu  e^{-ik_\nu(-b+R) } \rbrack,
\eea
where the prime stands for the differentiation with respect to the argument of the Airy functions,
 and from the normalization condition
\bea\label{nc}
  1 &=& \sum_{i=I}^{V} \int_{I_i} |\phi_i|^2 dx.
\eea 
Because of the relations
\bea
f_{II}(-b-R )&=& \gamma R_\nu \sim V_\infty^{1/3},
\nn
f_{II}(-b+R) &=&-\gamma r_\nu \sim -V_\infty^{-2/3}
\eea
with $R_\nu = 2R - r_\nu>0$ and $r_\nu= 2R\epsilon_\nu/V_\infty>0$, the arguments $f_{II}(-b-R )$ and
 $f_{II}(-b+R)$ take asymptotically large positive and small negative values, respectively, for
$V_\infty$ tending to infinity, for any $R\ge a$. Therefore we can use
the asymptotic formulas 10.4.59, 10.4.61, 10.4.63, 10.4.66 in Ref. \cite{Abraseg},
\bea
  {\rm{Ai}}(z) &\sim & \hf \pi^{-1/2} z^{-1/4} e^{-\zeta}\biggl(1+\ord{1/\zeta}\biggr),
~~~|{\rm{arg}}z|<\pi ,\nn
  {\rm{Ai}}^\prime(z)  &\sim & - \hf \pi^{-1/2} z^{+1/4} e^{-\zeta}\biggl(1+\ord{1/\zeta}\biggr),
~~~|{\rm{arg}}z|<\pi ,\nn
 {\rm{Bi}}(z) &\sim &   \pi^{-1/2} z^{-1/4} e^{+\zeta}\biggl(1+\ord{1/\zeta}\biggr),
~~~|{\rm{arg}}z|<\pi/3 ,\nn
  {\rm{Bi}}^\prime(z)  &\sim & \pi^{-1/2} z^{+1/4} e^{+\zeta}\biggl(1+\ord{1/\zeta}\biggr),
~~~|{\rm{arg}}z|<\pi/3 \nn
\eea
with  $z=\gamma R_\nu $ and $\zeta= \zeta_\nu=\frac{2}{3} (\gamma R_\nu)^{3/2}$ to obtain the asymptotic
 expressions
\bea\label{asyms}
  {\rm{Ai}}(f_{II}(-b-R)) &= &  {\rm{Ai}}(\gamma R_\nu) \nn
&\sim& \hf \pi^{-1/2} (\gamma R_\nu)^{-1/4} e^{-\zeta_\nu} 
 ,\nn
  {\rm{Bi}}(f_{II}(-b-R)) &= &  {\rm{Bi}}(\gamma R_\nu) \nn
&\sim& 
 \pi^{-1/2} (\gamma R_\nu)^{-1/4} e^{+\zeta_\nu}
,\nn
 {\rm{Ai}}^\prime(f_{II}(-b-R)) &= &  {\rm{Ai}}^\prime(\gamma R_\nu)\nn
&\sim& 
  - \hf \pi^{-1/2}(\gamma R_\nu)^{+1/4} e^{-\zeta_\nu},\nn
 {\rm{Bi}}^\prime(f_{II}(-b-R)) &= &  {\rm{Bi}}^\prime(\gamma R_\nu)\nn 
&\sim&
  \pi^{-1/2}(\gamma R_\nu)^{+1/4} e^{+\zeta_\nu }
\eea
for asymptotically large values of $V_\infty$ and for any $R\ge a$.
Furthermore, the expansions 10.4.2-10.4.5 in Ref. \cite{Abraseg},
\bea\label{zeroex}
 {\rm{Ai}}(z)&=& c_1 f(z) -c_2 g(z),\nn
{\rm{Bi}}(z) &=& \sqrt{3} \lbrack  c_1 f(z) +c_2 g(z) \rbrack,
\eea
with $c_1 = 3^{-2/3}/\Gamma(2/3)=0.355...$, $c_2=3^{-1/3}/\Gamma(1/3)=0.2588...$
and
\bea
  f(z)&=& 1+ \frac{z^3}{3!} + \ord{z^6},\nn
 g(z)&=& z+ \frac{2}{4!} z^4 + \ord{z^7}
\eea
with $z=-\gamma r_\nu$, $|z|\ll 1$ provide the behaviour of the Airy functions and their first derivatives in the neighbourhood of $z=0$, so that one finds
\bea\label{aiexpas}
   {\rm{Ai}}(f_{II}(-b+R)) &= &  {\rm{Ai}}(-\gamma r_\nu)
\approx c_1+c_2\gamma r_\nu,\nn
  {\rm{Bi}}(f_{II}(-b+R)) &= &  {\rm{Bi}}(-\gamma r_\nu)
\approx \sqrt{3}(c_1-c_2\gamma r_\nu),\nn
  {\rm{Ai}}^\prime (f_{II}(-b+R)) &= &  {\rm{Ai}}^\prime(-\gamma r_\nu)
\approx -c_2+ \frac{c_1}{2} (\gamma r_\nu)^2,\nn
 {\rm{Bi}}^\prime (f_{II}(-b+R)) &= &  {\rm{Bi}}^\prime(-\gamma r_\nu)
\approx  \sqrt{3}\biggl( c_2+ \frac{c_1}{2} (\gamma r_\nu)^2\biggr) .\nn
\eea
Making use of the expressions in Eqs. \eq{asyms} and \eq{aiexpas}, we can rewrite the boundary
 conditions in Eqs. \eq{bc1}-\eq{bc4} as
\bea\label{b1}
&& Ae^{-\kappa_\nu (b+R)} =
  \hf \pi^{-1/2} (\gamma R_\nu)^{-1/4} (  B e^{-\zeta_\nu}
+ 2C  e^{+\zeta_\nu}),\nn
&&\\
\label{b2}
&& A \kappa_\nu e^{-\kappa_\nu (b+R)}  \hf \pi^{-1/2}(\gamma R_\nu)^{+1/4}  \gamma ( B e^{-\zeta_\nu} - 2C  e^{+\zeta_\nu } ),\nn
&&\\
\label{b3}
&& B (c_1+c_2\gamma r_\nu)  + C \sqrt{3}(c_1-c_2\gamma r_\nu) \nn
&&~~~=
   D \lbrack e^{ik_\nu(-b+R) } + (-1)^\nu  e^{-ik_\nu(-2La+R) } \rbrack,
\\
\label{b4}
&&- \gamma\biggl\lbrack B\biggl( - c_2+ \frac{c_1}{2}(\gamma r_\nu)^2\biggr)
 + C \sqrt{3}\biggl( c_2+ \frac{c_1}{2}(\gamma r_\nu)^2 \biggr)
 \biggr\rbrack \nn
&&~~~ = ik_\nu  D \lbrack e^{ik_\nu(-b+R)  } - (-1)^\nu  e^{-ik_\nu(-b+R) } \rbrack .
\eea
Dividing the appropriate sides of Eqs. \eq{b1} and \eq{b2} and those of Eqs. \eq{b3} and \eq{b4},
we obtain a set of homogeneous linear equations for the determination of the coefficients 
$B$ and $C$,
\bea\label{BCrho}
 \rho_\nu = \frac{ Be^{-\zeta_\nu} - 2Ce^{\zeta_\nu}}{ 
  Be^{-\zeta_\nu} + 2C e^{\zeta_\nu} },~~~~~~~~~~~~~~~~~~~~~~~~~~~~~~~~~~~&&
\\
\label{BCxi}
   \Xi_\nu = \gamma \frac{ B \lbrack c_2-\frac{c_1}{2}(\gamma r_\nu)^2\rbrack
   - \sqrt{3}C \lbrack c_1+\frac{c_1}{2}(\gamma r_\nu)^2\rbrack }{ 
 B (c_1+c_2\gamma r_\nu)  +  \sqrt{3}C (c_1-c_2\gamma r_\nu)} ,~~&&
\eea
where $\rho_\nu$ and $\Xi_\nu$ on the left-hand sides are given as
\bea\label{rhonu}
\rho_\nu &=&\frac{\kappa_\nu  }{\gamma^{3/2}R_\nu^{1/2} }\nn
&= &
  \frac{\kappa}{\gamma^{3/2} (2R)^{1/2}}
 \biggl( 1 - \frac{\xi_\nu}{2} -\frac{\xi_\nu^2}{8}  \biggr)
\biggl( 1 + \frac{\xi_\nu}{2}  + \frac{3\xi_\nu^2}{8}\biggr)
\nn
&= & 
  1 +\ord{\xi_\nu^3}
\eea
with $0<\xi_\nu= \frac{r_\nu}{2R}\ll 1 $, $\kappa=  \sqrt{ 2mV_\infty}/\hbar$  and
\bea\label{xinu}
  \Xi_\nu &=&  ik_\nu
   \frac{ e^{ik_\nu(-b+R)  } - (-1)^\nu  e^{-ik_\nu(-b+R) } }{
      e^{ik_\nu(-b+R) } + (-1)^\nu  e^{-ik_\nu(-b+R) }      } \nn
&=&
  k_\nu 
\pmatrix{\tan (k_\nu(b-R) )  & \nu={\rm{even}} \cr
-{\rm{cotan}}  (k_\nu(b-R) ) &\nu={\rm{odd}} },
\eea
respectively. Eq. \eq{BCrho} with unity on the left-hand side in the limit $V_\infty\to \infty$
(c.f. Eq. \eq{rhonu}) can only be satisfied for $C=0$. Otherwise $ -1$ were the limiting
value of the right-hand side in contradiction with the limiting value $+1$ of the left-hand side.
 Then Eq. \eq{BCxi}  reduces to
\bea
   \Xi_\nu &=& \gamma \frac{   c_2-\frac{c_1}{2}(\gamma r_\nu)^2
   }{  c_1+c_2\gamma r_\nu } \sim \gamma \frac{c_2}{c_1}
\eea
that goes to $+\infty$ for $V_\infty\to \infty$. This implies, however, via the singularities of
 the tangent and cotangent functions in the right-hand side of Eq. \eq{xinu} that $k_\nu$ can be 
given as $k_\nu =k_\nu^\infty-\delta k_\nu$ with
\bea\label{kinfty}
  k_\nu^\infty   =\frac{\pi(\nu+1)}{ \ell-2R}
\eea
and the yet undetermined $\delta k_\nu$ for which the inequalities $0< \delta k_\nu\ll k_\nu^\infty$
hold for asymptotically large values of $V_\infty$. Making use of $C=0$ and  Eqs. \eq{kinfty} and \eq{xinu},
we can recast Eq. \eq{BCxi} in the form
\bea
  \frac{1}{\gamma} \frac{ c_1+c_2\gamma r_\nu }{ c_2-\frac{c_1}{2}(\gamma r_\nu)^2 }
&=& \frac{1}{k_\nu}
\pmatrix{{\rm{cotan}} (k_\nu(b-R) )  & \nu={\rm{even}} \cr
-\tan  (k_\nu(b-R) ) &\nu={\rm{odd}} }\nn
\eea
and expand the left-hand side in powers of $r_\nu$ and the right-hand side in powers of
 $\delta k_\nu/k_\nu^\infty$ (at the zeroth of the trigonometric functions). We keep the terms up to
 the first order in $\delta k_\nu/k_\nu^\infty$, while in the other expansion we have kept the terms 
of second order in  $r_\nu$ because in that manner we keep all terms of the order $a$ and $a^2$ 
until the limit $V_\infty\to \infty$ is taken. Then we find 
\bea
 \frac{\delta k_\nu}{k_\nu^\infty}&\approx &
 \frac{1}{\gamma (b-R)} \biggl( \frac{c_1}{c_2} +\gamma r_\nu
+ \frac{c_1^2}{2c_2^2}(\gamma r_\nu)^2 +\ord{r_\nu^3}\biggr).\nn
 \eea
The first term in the bracket provides the leading order term for asymptotically large $V_\infty$ and
$\delta k_\nu\to 0$ in the limit $V_\infty\to \infty$.

Now we have $C=0$ and the  coefficients $A$ and $D$ can be expressed from Eqs. \eq{b1} and 
\eq{b3} in terms of $B$,
\bea\label{AD}
 Ae^{-\kappa_\nu (b+R)} 
&=&
  \hf \pi^{-1/2} (\gamma R_\nu)^{-1/4}  e^{-\frac{2}{3}(\gamma R_\nu)^{3/2}} B\nn
&\approx &
  \hf \pi^{-1/2}(\gamma 2R)^{-1/4}  e^{-\zeta} 
\biggl(  1+ \frac{\xi_\nu}{4}(1+6\zeta)\nn
&&
+ \frac{\xi_\nu^2}{32} (  5 +36\zeta^2 )   +\ord{\xi^3}\biggr) B,\\
  D&=&
  \frac{ c_1+c_2\gamma r_\nu }{ 
    e^{ik_\nu(-b+R) } + (-1)^\nu  e^{-ik_\nu(-b+R) }  } B\nn
&\approx &
   \frac{ (-i)^\nu\gamma}{ 2k_\nu^\infty} 
\biggl( c_2 -\frac{c_1}{2} (\gamma r_\nu)^2+\ord{r_\nu^3}\biggr) B,
\nn
\eea
where the terms up to the quadratic ones in $\xi_\nu $ and $r_\nu$ have been kept and
$\zeta= \frac{2}{3}(\gamma 2R)^{3/2} $.

As to the next, one has to determine $B$ from the normalization condition \eq{nc} that can
 be recasted into the following form,
\bea\label{nc2}
  \hf |B|^{-2} &=& 
|Ae^{-\kappa_\nu (b+R)}/B|^2 \c{I}_A+ \c{I}_B
+ |D/B|^2 \c{I}_D\nn
\eea
with the integrals 
\bea\label{intabd}
 \c{I}_A &=&\int_{-\infty}^{0} dx e^{2\kappa  x}= \frac{1}{2\kappa_\nu}\nn
&\sim & \frac{1}{2\kappa} \biggl( 1+ \frac{\xi_\nu}{2} +\frac{3\xi_\nu^2}{8} +\ord{\xi_\nu^3}\biggr) ,\nn 
\c{I}_B&=& \int_{-2R+r_\nu}^{r_\nu} dx{\rm{Ai}}^2 (-\gamma x),\nn
\c{I}_D&=&\int_{-b+R}^0 dx | e^{ik_\nu x} + (-1)^\nu e^{-ik_\nu x} |^2\nn
&=&  \biggl(  \ell-2R +  \frac{(-1)^\nu}{k_\nu} \sin (k_\nu(\ell-2R))\biggr)\nn
&\sim &
\biggl( 1+ \frac{\delta k_\nu}{k_\nu^\infty}\biggr) (\ell-2R)
\eea
for asymptotically large values of $ V_\infty$. Since $0<r_\nu\ll R$ for
 sufficiently large $V_\infty$ and for any $R\ge a$,  we can estimate the integral $\c{I}_B$ by
 expanding it in powers of $r_\nu$,
\bea
\c{I}_B&\approx &
 \int_{-2R}^0  dx{\rm{Ai}}^2 (-\gamma x)
+r_\nu \lbrack {\rm{Ai}}^2 (0)-  {\rm{Ai}}^2 (\gamma 2R)\rbrack\nn
&&
 -  \gamma r_\nu^2 \lbrack  {\rm{Ai}} (0){\rm{Ai}}^\prime (0)
  -   {\rm{Ai}} (\gamma 2R){\rm{Ai}}^\prime (\gamma 2R) \rbrack .
\eea
Here the integral in the right-hand side is bounded in the limit $V_\infty\to \infty$,
  \bea
 \int_{-2R}^0  dx{\rm{Ai}}^2 (-\gamma x) &=&
  \int^{\gamma 2R}_0 dx{\rm{Ai}}^2 (x) \nn
& < &\int_0^\infty dx{\rm{Ai}}^2 (x)\equiv M_1,  
\eea
where $M_1>0$ is a constant independent of $\gamma$. Making use of the expansion
of the Airy function ${\rm{Ai}}(z)$ and  that of its first derivative
 at $z=0$  given in Eqs. \eq{zeroex} as well as their asymptotic expressions
 in Eqs. \eq{asyms} we find then the estimate
\bea\label{IB}
\c{I}_B&\approx &
M_1 +r_\nu \biggl(c_1^2 -\frac{1}{4\pi}   (\gamma 2R)^{-1/2}  e^{-2\zeta }
\biggr)\nn
&&
  -  \gamma r_\nu^2  \biggl( 
 -c_1c_2 + \frac{1}{4\pi}  e^{-2\zeta } \biggr).
\eea
Making use of Eqs. \eq{AD},  \eq{intabd} for the integrals $\c{I}_A$ and $\c{I}_D$, and the 
estimate  \eq{IB} of the integral $\c{I}_B$, we can rewrite  Eq. \eq{nc2} into the more explicit
 form
\bea
\lefteqn{
   \hf |B|^{-2}    }\nn &=& 
 \frac{  e^{-2\zeta}  }{8\pi\kappa \sqrt{\gamma 2R}}  
\biggl\lbrack 1 + \frac{3\xi_\nu}{4}  (1+2\zeta) +\frac{\xi_\nu^2}{32}( 21+24\zeta+ 36\zeta^2)
\biggr\rbrack\nn
&&
+M_1 +r_\nu \biggl(c_1^2 -\frac{  e^{-2\zeta } }{4\pi\sqrt{\gamma 2R}} 
\biggr)
  -  \gamma r_\nu^2  \biggl( 
 -c_1c_2 + \frac{  e^{-2\zeta}  }{4\pi }  \biggr)\nn
&&
+  \frac{\gamma^2 c_2^2}{ 2(k_\nu^\infty)^2}  \biggl\lbrack 
   2La-R + \frac{c_1}{\gamma c_2} + r_\nu
\nn
&&~~~~~~~~~~~~~~ -\frac{c_1}{c_2} \biggl(  2La-R + \frac{c_1}{2\gamma c_2}\biggr) (\gamma r_\nu)^2 
\biggr\rbrack.
\eea
Here $e^{-\zeta}$, $\gamma r_\nu$ and $\gamma^2 r_\nu$ vanish in the limit $V_0\to \infty$, therefore
 the only terms surviving that limit yield
\bea
  |B|^2 &\sim & \hf \biggl(  \frac{\gamma^2 c_2^2 (b-R)}{ 2(k_\nu^\infty)^2}
   + \frac{\gamma c_1c_2 }{ 2(k_\nu^\infty)^2} + M_1 \biggr)^{-1}  \nn
&\sim&
  \frac{ (k_\nu^\infty)^2  }{ \gamma^2 c_2^2 (b-R) } +\ord{1/\gamma^3}
\eea
in the leading order of $1/\gamma $ tending to zero.

Up to an irrelevant overall phase factor the   asymptotic values of the
 coefficients of the normalized wavefunctions are therefore
\bea
  B &\sim &  \frac{ k_\nu^\infty  }{ \gamma c_2 \sqrt{b-R} } ,
\eea
and from the expressions  in \eq{AD},
\bea\label{coefA}
   Ae^{-\kappa_\nu (b+R)} &\sim & 
\frac{ k_\nu^\infty e^{-\zeta}}{ 2c_2\gamma \sqrt{\pi (b-R)} (\gamma2R)^{1/4}},\\ 
%
\label{coefD}
 D&\sim &
  \frac{ (-i)^\nu}{ 2 \sqrt{b-R} }.
\eea
The asymptotic expressions of the pieces of the wavefunction of the stationary
 state $\nu$ in the trapezoid-well potential with asymptotically large depth
 are then given by the Eqs. in \eq{finupieces} and 
 by the parity relations $\phi_\nu(x)= (-1)^\nu \phi_\nu(-x)$, when we do not make the expression
 of the coefficient $B$ explicit.

\section{Reminder on nondegenerate stationary perturbation expansion}\label{perexp} 

Let us insert the perturbations expansions in Eqs. \eq{perham} and \eq{expas} into the stationary 
Schr\"odinger equation \eq{stsch} and write down the equalities of the terms on both sides of the
 equation order by order. The equation for the terms of zeroth order is just the stationary
 Schr\"odinger equation \eq{stschup} of the unperturbed system.  
 In the first and second orders of the perturbation expansion we find
\bea\label{ord1}
  \h{H}_0 \delta^{[1]} \psi_\nu +  \delta^{[1]}\h{H} \phi_\nu
&=&\epsilon_\nu \delta^{[1]} \psi_\nu +  \delta^{[1]}\epsilon_\nu \phi_\nu,
\eea
\bea
\label{ord2}
    \h{H}_0 \delta^{[2]} \psi_\nu +\delta^{[2]}\h{H} \phi_\nu +  \delta^{[1]}\h{H} \delta^{[1]} \psi_\nu 
&=&
 \epsilon_\nu \delta^{[2]} \psi_\nu +  \delta^{[2]}\epsilon_\nu \phi_\nu \nn
&&+  \delta^{[1]}\epsilon_\nu \delta^{[1]} \psi_\nu,
\eea
respectively. 

Both the first- $(n=1)$ and second-order $(n=2)$ corrections to the energy and the wavefunctions
 can be obtained by
 taking the scalar product of both sides of Eqs. \eq{ord1} and \eq{ord2}, respectively
 with the wavefunction $\phi_\mu$ and expanding the perturbative corrections
$ \delta^{[n]}\psi_\nu$ of the wavefunction in terms of the unperturbed wavefunctions $\phi_\mu$,
\bea
  \delta^{[n]}\psi_\nu &=& c_{\nu\nu}^{[n]}\phi_\nu + \sum_{\mu\not= \nu} c_{\nu\mu}^{[n]}\phi_\mu.
\eea
Then the case with $\mu=\nu$ yields the real energy shifts given in Eqs. \eq{epsfirst} and
 \eq{epssec} (remind that all operators $\delta^{[n]} \h{H}$ are Hermitian symmetric by construction)
and the case with $\mu\not=\nu$ provides the coefficients
\bea
  c_{\nu\mu}^{[1]}&\equiv& \la \phi_\mu| \delta^{[1]} \psi_\nu\ra
= -\frac{ \la \phi_\mu | \delta^{[1]}\h{H}| \phi_\nu \ra}{ \epsilon_\mu-\epsilon_\nu}
\eea
and
\bea
   c^{[2]}_{\nu\mu} &=&
- \frac{ \la \phi_\mu |\delta^{[2]}H |\phi_\nu\ra }{ \epsilon_\mu
-\epsilon_\nu }\nn
&&
 +  \sum_{\mu'\not=\nu} \frac{ \la \phi_\mu| \delta^{[1]}H| \phi_{\mu'} \ra
 \la \phi_\mu' | \delta^{[1]}H| \phi_\nu\ra }{(\epsilon_\mu-\epsilon_\nu)
 (\epsilon_{\mu'}-\epsilon_\nu ) }\nn
&&
-\frac{  \la \phi_\nu |\delta^{[1]}H|\phi_\nu\ra\la \phi_\mu | \delta^{[1]}H| \phi_\nu\ra
}{
(\epsilon_\mu-\epsilon_\nu)^2 }.
\eea
The coefficients $c^{[1]}_{\nu\nu}$ and $c^{[2]}_{\nu\nu}$  should be determined from the
  normalization to unity of the first- and second-order improved wavefunctions, respectively,
with the appropriate accuracy. Then one finds $c^{[1]}_{\nu\nu}=0$ and
\bea
  c^{[2]}_{\nu\nu} &=& -\hf \sum_{\mu\not= \nu} \frac{ |\la \phi_\mu|\delta^{[1]}\h{H} |
\phi_\nu\ra |^2 }{ (\epsilon_\mu-\epsilon_\nu)^2 }
\eea
and obtains the expressions in Eqs. \eq{psifirst} and \eq{psisec} finally.

\section{Evaluation of  matrix elements}\label{matele}

In this section we outline the evaluation of the integrals
 $(v^{[1]})^{II,III}_{\nu'\nu}$, $(v^{[1]})^{II,I}_{\nu'\nu}$, $(v^{[1]})^{II,II}_{\nu'\nu}$
, $(t^{[1]})^{III,III}_{\nu'\nu}$, $(t^{[1]})^{II,III}_{\nu'\nu}$,  $(t^{[1]})^{II,II}_{\nu'\nu}$,  
$(t^{[1]})^{II,I}_{\nu'\nu}$,  $(h_t^{[2]}+t^{[2]})^{III,III}_{\nu'\nu}$,  $(h_t^{[2]}+t^{[2]})^{II,III}_{\nu'\nu}$,
 $(h_t^{[2]}+t^{[2]})^{II,II}_{\nu'\nu}$, and $(h_t^{[2]}+t^{[2]})^{II,I}_{\nu'\nu}$ for  finite but 
asymptotically large values of $V_\infty$. These are the integrals contributing to the matrix elements 
which may become nonvanishing in the limit $V_\infty\to \infty$. Evaluating the integrals $K^{ij}_{\nu'\nu}$
 according to Eq. \eq{kijs} one has to keep in mind that (i) the term containing the Dirac delta does
 not contribute if the intervals $I_i$ and $I_j$ are different, and (ii) the wavefunctions are real. One 
obtains the explicit expressions for the integrals  $K^{ij}_{\nu'\nu}$  for asymptotically large values 
of the parameter $V_\infty$  by inserting the expressions \eq{v1}, \eq{t1}, \eq{ht2}, and \eq{t2} of the
 appropriate  kernels and the expressions of the appropriate pieces of the wavefunctions  given in Eq.
 \eq{finupieces} into the double integral in Eq. \eq{kijs}.  In order to make our formulas more compact
let us use the notations
\bea
 \int_{I}dx f(x)&=& \int_{-\infty}^{-b-R} dx f(x), \nn
 \int_{II} dx f(x)&=& \int_{-b-R}^{-b+R} dx f(x), \nn
 \int_{III} dx f(x)&=& \int_{-b+R}^{b-R} dx f(x)
\eea
with $b=\ell/2$.

\subsection{Determination of $(v^{[1]})_{\nu'\nu}^{ij}$}

Let us start with the integral $(v^{[1]})^{II,III}_{\nu'\nu}$ that reduces to
\bea\label{v1_II_III}
\lefteqn{
 (v^{[1]})^{II,III}_{\nu'\nu}    }\nn
&=& 
\int_{II} dx \int_{III} dy  \phi_{\nu' II}(x)^* \hf \lbrack
 V_{t}(x)+V_{t}(y) \rbrack\nn
&&\times \Pi(x-y)\phi_{\nu III}(y)\nn
&\sim &
\frac{V_\infty}{4R}  \frac{ k_{\nu'}^\infty  (-i)^\nu }{2 \gamma c_2 (b-R) } \nn
&&\times
 \int_{II} dx \int_{III} dy 
 {\rm{Ai}}( \gamma (-x -(b-R) -r_{\nu'}))\nn
&&\times  \lbrack (R-x-b)+ (R-y-b)\rbrack \Pi(x-y)\nn
&&\times
 \lbrack e^{ik_\nu y} +(-1)^\nu e^{-ik_\nu y}\rbrack.
\eea
Let us make use of the inequality $2R\ll \ell$ and expand the integral  over the variable $x$
as the function of its limits in the small parameter $2R/\ell$, keeping the leading order term,
\bea
\lefteqn{
 (v^{[1]})^{II,III}_{\nu'\nu}             }\nn
&\sim &
\frac{V_\infty}{4R}   \frac{ k_{\nu'}^\infty  (-i)^\nu }{2 \gamma c_2 (b-R) }    2R
 {\rm{Ai}}( \gamma (R -r_{\nu'}))\nn
&&\times\int_{III} dy  \lbrack R + (R-y-2La)\rbrack     \Pi(-2La-y)\nn
&&\times
 \lbrack e^{ik_\nu y} +(-1)^\nu e^{-ik_\nu y}\rbrack.
\eea
The remaining integral over the variable $y$ has finite limit, while the argument of the Airy
 function takes asymptotically large values for $V_\infty\to \infty$ and therefore the
 Airy-function suppresses exponentially the integral $ (v^{[1]})^{II,III}_{\nu'\nu} \to 0$ in
 that limit.  

Replacing the function $\phi_{\nu III}(y)$ by $\phi_{\nu I}(y)$ and changing the integral 
$\int_{III} dy$ to $\int_{I} dy$ in Eq. \eq{v1_II_III}, we find in a similar manner that
$ (v^{[1]})^{II,I}_{\nu'\nu}\to 0$ in the limit $V_\infty\to \infty$.

The integral $(v^{[1]})^{II,II}_{\nu'\nu}$ can be rewritten as
\bea
\lefteqn{
 (v^{[1]})^{II,II}_{\nu'\nu}   }\nn
&= &
\frac{V_\infty}{4R} \frac{ k_{\nu'}^\infty k_\nu^\infty  }{ \gamma^2 c_2^2 (b-R) }\nn
&&\times
  \int_{II} dx\int_{II}  dy
 {\rm{Ai}}( \gamma (-x -b+R -r_{\nu'}))\nn
&&\times
\lbrack -\ell + 2R -x-y\rbrack\lbrack \Pi(x-y)-\delta (x-y)\rbrack\nn
&&\times
   {\rm{Ai}}( \gamma (-y -(b-R) -r_{\nu})). 
\eea
Let us first expand the integral over $y$ as the functions of its limits in the small parameter
 $2R/\ell$, keeping the leading order terms, and then do the same for the $x$-integral. Then one finds
\bea
(v^{[1]})^{II,II}_{\nu'\nu} 
&\sim &
 \frac{V_\infty}{4R} \frac{ k_{\nu'}^\infty k_\nu^\infty  }{ \gamma^2 c_2^2 (b-R) }
(2R){\rm{Ai}}( \gamma (R -r_{\nu})) \nn
&&\times
\lbrack  (2R){\rm{Ai}}( \gamma ( R -r_{\nu'})) ( 2R)  \Pi(0)\nn
&&~~~
- 2R {\rm{Ai}}( \gamma (R -r_{\nu'})) \rbrack.
\eea
Because of the Airy functions with arguments tending to plus infinity this integral is exponentially
 suppressed in the limit $V_\infty\to \infty$.

\subsection{Determination of $(t^{[1]})_{\nu'\nu}^{ij}$ and $(h_t^{[2]}+t^{[2]})^{ij}_{\nu'\nu}$}

Let us evaluate now the integrals $K^{ij}_{\nu'\nu}$ connected with the projection of the kinetic 
energy operator. Let us start with the integral  $(t^{[1]})^{III,III}_{\nu'\nu}$ that can be
rewritten as the sum of two integrals, $(t^{[1]})^{III,III}_{\nu'\nu} =(t^{[1]}_\Pi)^{III,III}_{\nu'\nu} 
+(t^{[1]}_D)^{III,III}_{\nu'\nu}$ with the terms containing the projector and the Dirac-delta,  
\bea
\lefteqn{
(t^{[1]}_\Pi)^{III,III}_{\nu'\nu}       }\nn
&= &
-  \frac{\hbar^2}{2m} \frac{ i^{\nu'-\nu} }{4(b-R)}
 \int_{III} dx \int_{III} dy 
\lbrack e^{-ik_{\nu'} x} \nn
&&+(-1)^{\nu'} e^{ik_{\nu'} x}\rbrack
\partial_x^2  \Pi(x-y) \lbrack e^{ik_\nu y} +(-1)^\nu e^{-ik_\nu y}\rbrack,\nn
\eea
and
\bea\label{ii1}
\lefteqn{
(t^{[1]}_D)^{III,III}_{\nu'\nu}   }\nn
&= &
-  \frac{\hbar^2 k_\nu^2 }{2m} \frac{ i^{\nu'-\nu} }{4(b-R)}
  \int_{III} dx
\lbrack e^{-ik_{\nu'} x} \nn 
&&+(-1)^{\nu'} e^{ik_{\nu'} x}\rbrack
  \lbrack e^{ik_\nu x} +(-1)^\nu e^{-ik_\nu x}\rbrack,
\eea
respectively. The integral in the term  $(t^{[1]}_D)^{III,III}_{\nu'\nu}$ can be performed easily 
and yields
\bea
\lefteqn{
 (t^{[1]}_D)^{III,III}_{\nu'\nu} }\nn
&= &-  \frac{\hbar^2 k_\nu^2 }{2m} \frac{ i^{\nu'-\nu} \lbrack  1   + (-1)^{\nu'+\nu}\rbrack  
   }{2(b-R)}\biggl(
  \frac{  \sin ( (k_{\nu'}-k_\nu) (b-R)) }{  k_{\nu'}-k_\nu }\nn
&&
+ (-1)^\nu \frac{  \sin( (k_{\nu'}+k_\nu) (b-R)) }{ k_{\nu'}+k_\nu } \biggr).
\eea
We can take now the limit $V_\infty\to \infty$ that implies  $k_\nu\to k_\nu^\infty$ and making 
use of Eq. \eq{kinfty}, one finds
\bea\label{ii}
(t^{[1]}_D)^{III,III}_{\nu'\nu}  &=&-  \frac{\hbar^2 (k_\nu^\infty)^2 }{2m} \delta _{\nu'\nu}.
\eea
It is straightforward to obtain this result for $\nu'=\nu$. For $\nu'\not=\nu$
one gets the expression
\bea
\lefteqn{
(t^{[1]}_D)^{III,III}_{\nu'\nu}       }\nn
&= &-  \frac{\hbar^2 (k_\nu^\infty)^2 }{2m}  i^{\nu'-\nu} \lbrack  1   + (-1)^{\nu'+\nu}\rbrack  
   \biggl(
  \frac{  \sin (\frac{\pi}{2} (\nu'-\nu) ) }{  \pi(\nu'-\nu) }\nn
&&
+ (-1)^\nu \frac{  \sin( \frac{\pi}{2} (\nu'+\nu+2) ) }{ \pi(\nu'+\nu+2) } \biggr)
\eea
that vanishes if both $\nu'$ and $\nu$ are even or odd (i.e., $\nu'-\nu$ and $\nu'+\nu$ are even)
 due to the vanishing of the values of the  sine functions, and it vanishes also when one of 
the integers $\nu'$ and $\nu$ is even and the other odd (i.e., $\nu'-\nu$ is odd) due to the
factor in the square bracket. Thus one finds the expression in Eq. \eq{ii}, that
equals to the negative of the unperturbed kinetic energy of the particle in the cavity for
 $\nu'=\nu$ and vanishes for $\nu'\not= \nu$. Let us note that the expressions obtained in the limit 
$V_\infty\to \infty$ are independent of the ratio $(\ell-2R)/a$.

The evaluation of the  term $(t^{[1]}_\Pi)^{III,III}_{\nu'\nu}  $
is more cumbersome. In order to estimate this integral, we perform its evaluation similarly to that done
 for $t_{II,II}$ in Eq. (E8) in \cite{Sail2013}. First we rewrite the integral $\int_{-d}^d dy \delta (x-y) 
e^{iky}=e^{ikx} \chi_{\lbrack -d,d\rbrack }(x)$, where $ \chi_{\lbrack -d,d\rbrack }(x)$ is the characteristic
 function of the interval $x\in \lbrack-d,d\rbrack $, as the limit
\bea
  \int_{-d}^d dy  \delta (x-y) e^{iky} &=& e^{ikx} \lim_{\Lambda\to \infty} \c{I}_\Lambda (x)
\eea
with 
\bea 
\c{I}_\Lambda (x) &=& \hf \int_{-1}^1 ds \biggl\lbrack \frac{\sin \lbrack s\Lambda (x+d)\rbrack
}{s\pi}  -  \frac{\sin \lbrack s\Lambda (x-d)\rbrack
}{s\pi} \biggr\rbrack .\nn
\eea
Then we realize that for $x\in \lbrack -d,d\rbrack$ the integral $\int_{-d}^d dy \Pi(x-y)
e^{iky}$ can be rewritten as
\bea
   \int_{-d}^d dy \Pi(x-y)e^{iky}&=& e^{ikx}\c{I}_K (x)
\eea
Below we have to set $d=b-R$. The cutoff $\Lambda$ has been replaced by $K$ and the limit removed.
 For  extremely large values of  $K=\pi/a$ the function $\c{I}_K (x)$ should be rather smooth, because
 in the limit  $K\to \infty$ it tends to 1, a value independent of $x$. Therefore, we can write as a good
 approximation
\bea\label{derofpi}
  \partial_x^2  \int_{-d}^d dy \Pi(x-y)
e^{iky}&\approx & -k^2  e^{ikx}\c{I}_K (x)
\eea
for $x\in \lbrack -d,d\rbrack$, neglecting the terms with the derivatives of $\c{I}_K(x)$. Now the
 first integral term of $(t^{[1]})^{III,III}_{\nu'\nu}$ can be estimated as
\bea\label{i1}
(t^{[1]}_\Pi)^{III,III}_{\nu'\nu}
&\approx &
 \frac{\hbar^2k_{\nu}^2}{2m} \frac{i^{\nu'-\nu} }{4(b-R)} \c{I}_K(0)
 \int_{-b+R}^{b-R} dx\lbrack e^{-ik_{\nu'} x} \nn
&&+(-1)^{\nu'} e^{ik_{\nu'} x}\rbrack
 \lbrack e^{ik_\nu x} +(-1)^\nu e^{-ik_\nu x}\rbrack ,\nn
\eea
where we replaced the very slowly  varying function $\c{I}_K(x)$ by its value at $x=0$,
in the middle of the potential well,
\bea
  \c{I}_K(0) &=& \int_{-1}^1 ds\frac{\sin (sK(b-R))}{ s\pi}
= \frac{2}{\pi}{\rm{Si}}\lbrack K(b-R) \rbrack .\nn
\eea
Here ${\rm{Si}}(u)$ denotes  the sine integral function. The integral standing in Eq. \eq{i1}  
is just the same as the one in Eq. \eq{ii1} vanishing for $\nu'\not=\nu$ and taking the value 
$4(b-R)$ for $\nu'=\nu$ in the limit $V_\infty\to \infty$, so that we obtain
\bea
(t^{[1]}_\Pi)^{III,III}_{\nu'\nu}   &=&  \frac{\hbar^2 (k_{\nu}^\infty)^2}{2m} \c{I}_K(0)
 \delta _{\nu'\nu}
\eea
and
\bea
  (t^{[1]})^{III,III}_{\nu'\nu} &=&  
 \frac{\hbar^2 (k_{\nu}^\infty)^2}{2m} \lbrack \c{I}_K(0)-1\rbrack \delta _{\nu'\nu}.
\eea
The corrections of various orders in the small parameter $a/\ell$ can be made explicit
by  making use of the asymptotic expansion of the sine integral function (see
the relations 5.2.6, 5.2.7, 5.2.8, 5.2.34, 5.2.35 in \cite{Abraseg}). Keeping the terms up
 to the order $(a/\ell)^2$, we find
\bea\label{asymp}
\c{I}_K(0)-1
&\sim &
 - \frac{2}{\pi} \biggl(\frac{ a \cos (K(b-R))}{\pi(b-R) } 
  -  \frac{a^2\sin (K(b-R)) }{ \pi^2(b-R)^2}  \biggr).\nn
\eea

Let us now consider the integral $(t^{[1]})^{II,III}_{\nu'\nu}$. The term of the kernel
with the Dirac-delta does not contribute and the integral reduces to
\bea 
\lefteqn{
(t^{[1]})^{II,III}_{\nu'\nu} }\nn
&=& -\frac{\hbar^2}{2m}
\frac{ k_{\nu'}^\infty (-i)^\nu}{2\gamma c_2(b-R)}  \nn
&&\times
\int_{II}dx \int_{III} dy {\rm{Ai}}(\gamma(-x-(-b-R)-r_{\nu'}))\nn
&&\times \partial_x^2\Pi(x-y)(e^{ik_{\nu}y}+(-1)^{\nu}e^{-ik_{\nu}y}).
\eea
Let us  apply the approximation given in Eq. \eq{derofpi} and make the replacement
$\c{I}_K(x)\Rightarrow \c{I}_K(0)$, again. Then we get
\bea
(t^{[1]})^{II,III}_{\nu'\nu}
&\approx& \frac{\hbar^2k_\nu^2}{2m}
\frac{ k_{\nu'}^\infty (-i)^\nu}{2\gamma c_2(b-R)}\c{I}_K(0)  \nn
&&\times
\int_{II}dx {\rm{Ai}}(\gamma(-x-(-b-R)-r_{\nu'}))\nn
&&\times
\lbrack e^{ik_{\nu}x}+(-1)^{\nu}e^{-ik_{\nu}x}\rbrack.
\eea
As to the next we expand the integral $\int_{II} dx \ldots$ as the function of its limits in
the small parameter $2R/\ell$, keeping the leading order term,
\bea
\lefteqn{
(t^{[1]})^{II,III}_{\nu'\nu}   }\nn
&=& 
\frac{\hbar^2k_\nu^2}{2m}
\frac{ k_{\nu'}^\infty R (-i)^\nu}{\gamma c_2(b-R)}\c{I}_K(0)  
{\rm{Ai}}(\gamma(\ell +R-r_{\nu'})\nn
&&\times
\lbrack e^{-ik_{\nu}b}+(-1)^{\nu}e^{ik_{\nu}b}\rbrack.
\eea
In the limit $V_\infty\to \infty$ the argument of the Airy function goes to plus infinity, too.
Therefore, the integral $(t^{[1]})^{II,III}_{\nu'\nu} $ is exponentially suppressed
in the limit $V_\infty\to \infty$.

In order to estimate the integral $(t^{[1]})^{II,II}_{\nu'\nu}$ we split it into the sum
of the terms with the projector and the Dirac-delta, $(t^{[1]})^{II,II}_{\nu'\nu}=
(t^{[1]}_\Pi)^{II,II}_{\nu'\nu}+(t^{[1]}_D)^{II,II}_{\nu'\nu}$, where
\bea
\label{t1i}
(t^{[1]}_\Pi)^{II,II}_{\nu'\nu}  &=&
-\frac{\hbar^2}{2m} \frac{k_{\nu'}^\infty k_\nu^\infty }{\gamma^2 c_2^2 (b-R)}\nn
&&\times \int_{II} dx \int_{II} dy {\rm{Ai}}(\gamma(-x-(2La-R)-r_{\nu'}))\nn
&&\times \partial_x^2 \Pi(x-y)
 {\rm{Ai}}(\gamma(-y-(b-R)-r_{\nu})),\nn
&&\\ 
\label{t1ii} 
(t^{[1]}_D)^{II,II}_{\nu'\nu} &=&
\frac{\hbar^2}{2m} \frac{k_{\nu'}^\infty k_\nu^\infty }{\gamma^2 c_2^2 (b-R)}\nn
&&\times \int_{II} dx \int_{II} dy {\rm{Ai}}(\gamma(-x-(b-R)-r_{\nu'}))\nn
&&\times \partial_x^2\delta (x-y)\rbrack
 {\rm{Ai}}(\gamma(-y-(b-R)-r_{\nu})) \nn
&=&
\frac{\hbar^2}{2m} \frac{k_{\nu'}^\infty k_\nu^\infty }{\gamma^2 c_2^2 (b-R)}\nn
&&\times \int_{II} dx {\rm{Ai}}(\gamma(-x-(b-R)-r_{\nu'}))\nn
&&\times \partial_x^2 {\rm{Ai}}(\gamma(-x-(b-R)-r_{\nu})). 
\eea
In Eq. \eq{t1ii} the integral over $x$ as the function of its limits can be expanded
in the small parameter $2R/\ell$, and approximated by its leading order term as
\bea
 (t^{[1]}_D)^{II,II}_{\nu'\nu} &\approx&
\frac{\hbar^2}{2m} \frac{2k_{\nu'}^\infty k_\nu^\infty R }{\gamma^2 c_2^2 (b-R)}
 {\rm{Ai}}(\gamma(R-r_{\nu'}))\nn
&&\times
\gamma^2  {\rm{Ai}}^{\prime\prime}(\gamma(R-r_{\nu})).
\eea
Since  the arguments of the Airy function and its second derivative tend to
 $\gamma R\to +\infty$ in the limit $V_\infty\to \infty$, this term is exponentially suppressed
 and vanishes in that limit. Now let us turn to the term \eq{t1i}, insert the projector
in its integral form given in Eq. \eq{projec}  and  perform the differentiation $\partial_x^2$,
\bea
 (t^{[1]}_\Pi)^{II,II}_{\nu'\nu} &=&
\frac{\hbar^2}{2m} \frac{k_{\nu'}^\infty k_\nu^\infty }{\gamma^2 c_2^2 (b-R)}
\int_{-K}^K \frac{dk_x}{2\pi} k_x^2
\nn
&&\times \int_{II} dx \int_{II} dy {\rm{Ai}}(\gamma(-x-(b-R)-r_{\nu'}))\nn
&& e^{ik_x(x-y)}
 {\rm{Ai}}(\gamma(-y-(b-R)-r_{\nu})),
\eea
and then  replace $x$ and $y$ in the Airy function slowly varying in the narrow interval
$I_{II}$ by their values at $-b$, i.e., at the middle of the interval,
\bea\label{t1_2_2restin}
\lefteqn{
(t^{[1]}_\Pi)^{II,II}_{\nu'\nu}     }\nn &\approx &
\frac{\hbar^2}{2m} \frac{k_{\nu'}^\infty k_\nu^\infty }{\gamma^2 c_2^2 (b-R)}
 {\rm{Ai}}(\gamma(R-r_{\nu'})){\rm{Ai}}(\gamma(R-r_{\nu}))\nn
&&\times
\int_{-K}^K \frac{dk_x}{2\pi} k_x^2 \int_{II} dx \int_{II} dy  e^{ik_x(x-y)}.
\eea
The remaining integrals can be taken exactly and one finds
\bea\label{t1_2_2limit}
\lefteqn{
(t^{[1]}_\Pi)^{II,II}_{\nu'\nu}    }\nn
 &\approx & \frac{\hbar^2}{2m} \frac{k_{\nu'}^\infty k_\nu^\infty }{\gamma^2 c_2^2 (b-R)}
 {\rm{Ai}}(\gamma(R-r_{\nu'})) {\rm{Ai}}(\gamma(R-r_{\nu}))\nn
&&\times \frac{2}{a} \biggl( 1 -\frac{\sin (2KR)}{2KR} \biggr).
\eea
We see again that for large values of $V_\infty$ this contribution is exponentially suppressed 
due to the asymptotically large arguments of the Airy functions, so that it vanishes in the limit
 $V_\infty\to \infty$. It is worthwhile mentioning that if $V_\infty$ would take a finite but large
value, the estimate of  $(t^{[1]})^{II,II}_{\nu'\nu}$ would be of the order $\sim \gamma^{-2}a^{-1}
\sim a^{-1/3}$. This nonanalytic dependence of the integral on the minimal length scale
$a$ is a hint that the treatment of the indirect GUP effect as a perturbation may not work when 
the potential well is of finite depth.

Let us now turn to the integral $(t^{[1]})^{I,II}_{\nu'\nu}$ given as
\bea
 (t^{[1]})^{I,II}_{\nu'\nu}&=&
    -\frac{\hbar^2}{2m} \frac{ k_{\nu'}^\infty k_\nu^\infty e^{-\zeta} }{
   2\sqrt{\pi}c_2^2\gamma^2 (b-R)(\gamma 2R)^{1/4}   } \nn
&&\times
\int_I dx \int_{II} dy e^{\kappa_{\nu'}(x+b+R)} \partial_x^2\Pi(x-y)\nn
&&\times
 {\rm{Ai}}(\gamma(-y-(b-R)-r_{\nu})).
\eea
Making the same approximation as we did for the evaluation of $(t^{[1]}_\Pi)^{II,II}_{\nu'\nu}$,
we find the analogue of Eq. \eq{t1_2_2restin},
\bea\label{t1_1_2restin}
\lefteqn{
(t^{[1]})^{I,II}_{\nu'\nu}     }\nn &\approx &
\frac{\hbar^2}{2m}\frac{ k_{\nu'}^\infty k_\nu^\infty e^{-\zeta} }{
   2\sqrt{\pi}c_2^2\gamma^2 (b-R)(\gamma 2R)^{1/4}   }
{\rm{Ai}}(\gamma(R-r_{\nu}))\nn
&&\times
\int_{-K}^K \frac{dk_x}{2\pi} k_x^2 \int_{I} dx e^{\kappa_{\nu'}(x+b+R)}  \int_{II} dy  e^{ik_x(x-y)}.\nn
\eea
The remaining threefold integral  reduces to the integral
\bea
  \c{J}_K&=& \int_{-K}^K \frac{dk_x}{2\pi} 
\frac{k_x^2 \lbrack \cos (k_x2R) -1\rbrack + \kappa_{\nu'} k_x\sin(k_x 2R)}{
   \kappa_{\nu'}^2 +k_x^2}\nn
\eea
exhibiting a finite limit for $V_\infty\to \infty$. Then the integral $ (t^{[1]})^{I,II}_{\nu'\nu}$
vanishes in the limit $V_\infty\to\infty$ due to the factor $e^{-\zeta}$ and the asymptotically large
argument of the Airy function.  

The integrals   $( h^{[2]}_t+ t^{[2]})_{\nu'\nu}^{i,j}$ can be evaluated and estimated
in a rather similar manner as it was done for the integrals  $(t^{[1]})^{i,j}_{\nu'\nu}$.
Considering the sum   $( h^{[2]}_t+ t^{[2]})_{\nu'\nu}^{i,j}$ instead of the integrals 
 $( h^{[2]}_t)_{\nu'\nu}^{i,j}$  and $(  t^{[2]})_{\nu'\nu}^{i,j}$ separately has the advantage that
the sum contains only the projected kinetic energy kernel.  The estimates of the integrals
$( h^{[2]}_t+ t^{[2]})_{\nu'\nu}^{i,j}$ for $(i,j)=(III,III)$ and $(II,III)$ can be obtained
 from the corresponding estimates of the integrals $(t^{[1]}_\Pi)^{i,j}_{\nu'\nu}$
when  their factor $(\hbar k_\nu^\infty)^2/(2m)$ is multiplied with $(2/3)(\alpha\hbar k_\nu^\infty)^2$
 that corresponds to the replacement of the ordinary kinetic energy operator 
in the matrix elements of $\h{t}^{[1]}$ with
the operator $(2/3) \alpha^2 (-i\hbar \partial_x)^4 /(2m)$ in the matrix elements of
$\h{h}_t^{[2]}+\h{t}^{[2]}$. In that manner one finds that $( h^{[2]}_t+ t^{[2]})_{\nu'\nu}^{II,III}$
is exponentially suppressed  for large values of the parameter $V_\infty$ and vanishes
 in the limit $V_\infty\to \infty$, while one obtains
\bea\label{htt2_3_3}
  ( h^{[2]}_t+ t^{[2]})_{\nu'\nu}^{III,III} &\sim & \delta_{\nu'\nu} 
\frac{2\alpha^2}{3}\frac{\hbar^4 (k_\nu^\infty)^4}{2m}  \c{I}_K(0)
\eea
in the limit $V_\infty\to \infty$. The same replacement of the ordinary kinetic energy operator
in order to obtain  the integrals $( h^{[2]}_t+ t^{[2]})_{\nu'\nu}^{i,j}$ 
from the integrals $(t^{[1]}_\Pi)^{i,j}_{\nu'\nu}$ with $(i,j)=(II,II)$ and $(I,II)$
leads to the slight modification of the remaining  integrals over the wave vector $k_x$
in Eqs. \eq{t1_2_2restin} and \eq{t1_1_2restin}, namely to the replacement  $k_x^2 \Rightarrow
 {\rm{const.}}\times k_x^4$ in their integrands. This, however, does not alter the conclusion
 that the remaining integrals yield finite results, and that the integrals
  $( h^{[2]}_t+ t^{[2]})_{\nu'\nu}^{i,j}$ with  $(i,j)=(II,II)$ and $(I,II)$ vanish in the limit
$V_\infty\to \infty$. Thus we conclude that the only nonvanishing contribution to the matrix elements
$  ( h^{[2]}_t+ t^{[2]})_{\nu'\nu}$ is given by Eq. \eq{htt2_3_3}. The leading order term of the
factor $\c{I}_K(0)$ is independent of the minimal length scale $a$ and it holds
 $\c{I}_K(0)\approx 1+ \ord{a/\ell}$ according to the expansion given in Eq. \eq{asymp}.
 Therefore, the contribution in Eq.   \eq{htt2_3_3} is just due to the direct GUP effect modifying 
the kinetic energy operator explicitly, the effect of the restriction of the kinetic energy operator 
to the subspace of the bandlimited wavefunctions by projection, i.e., the indirect GUP effect results
 in terms of third and higher orders, to be neglected in the second order of the perturbation expansion.

\end{document}